\documentclass[sn-mathphys,Numbered]{Definitions/sn-jnl}% Math and Physical Sciences Reference Style
\usepackage{graphicx}%
\usepackage{multirow}%
\usepackage{amsmath,amssymb,amsfonts}%
\usepackage{amsthm}%
\usepackage{mathrsfs}%
\usepackage[title]{appendix}%
\usepackage{xcolor}%
\usepackage{textcomp}%
\usepackage{manyfoot}%
\usepackage{booktabs}%
\usepackage{algorithm}%
\usepackage{algpseudocode}%
\usepackage{listings}%
\usepackage{todonotes}
\usepackage{mathtools}
\usepackage{subcaption}
\usepackage{gensymb}

\begin{document}
	
	\title[Article Title]{A Reinforcement Learning Approach for Robust Supervisory Control of UAVs Under Disturbances}
	
	%%=============================================================%%
	%% Prefix	-> \pfx{Dr}
	%% GivenName	-> \fnm{Joergen W.}
	%% Particle	-> \spfx{van der} -> surname prefix
	%% FamilyName	-> \sur{Ploeg}
	%% Suffix	-> \sfx{IV}
	%% NatureName	-> \tanm{Poet Laureate} -> Title after name
	%% Degrees	-> \dgr{MSc, PhD}
	%% \author*[1,2]{\pfx{Dr} \fnm{Joergen W.} \spfx{van der} \sur{Ploeg} \sfx{IV} \tanm{Poet Laureate} 
	%%                 \dgr{MSc, PhD}}\email{iauthor@gmail.com}
	%%=============================================================%%
	
	\author[1]{\fnm{Ibrahim} \sur{Ahmed}}\email{ibrahim.ahmed@vanderbilt.edu}
	
	\author[2]{\fnm{Marcos} \sur{Quinones-Grueiro}}\email{marcos.quinones.grueiro@vanderbilt.edu}
	
	\author*[2]{\fnm{Gautam} \sur{Biswas}}\email{gautam.biswas@vanderbilt.edu}
	
	\affil[1]{\orgdiv{Electrical and Computer Engineering, Institute for Software Integrated Systems}, \orgname{Vanderbilt University}, \orgaddress{\street{1025 16th Ave S}, \city{Nashville}, \postcode{37212}, \state{TN}, \country{USA}}}
	
	\affil[2]{\orgdiv{ Computer Science, Institute for Software Integrated Systems}, \orgname{Vanderbilt University}, \orgaddress{\street{1025 16th Ave S}, \city{Nashville}, \postcode{37212}, \state{TN}, \country{USA}}}
	
	%%==================================%%
	%% sample for unstructured abstract %%
	%%==================================%%
	
	\abstract{In this work, we present an approach to supervisory reinforcement learning control for unmanned aerial vehicles (UAVs). UAVs are dynamic systems where control decisions in response to disturbances in the environment have to be made in the order of milliseconds. We formulate a supervisory control architecture that interleaves with extant embedded control and demonstrates robustness to environmental disturbances in the form of adverse wind conditions. We run case studies with a Tarot T-18 Octorotor to demonstrate the effectiveness of our approach and compare it against a classic  cascade control architecture used in most vehicles. While the results show the performance difference is marginal for nominal operations, substantial performance improvement is obtained with the supervisory RL approach under unseen wind conditions.}
	
	
	%%\pacs[JEL Classification]{D8, H51}
	
	%%\pacs[MSC Classification]{35A01, 65L10, 65L12, 65L20, 65L70}
	
	\maketitle
	
	\section{Introduction}\label{sec:intro}
	
	Advances in technology %in battery and materials 
	have increased the feasibility of deploying UAVs across many different applications that include transportation, surveillance, nature research, extreme sports,  professional photography, and surveying \citep{cohn2017commercial,iv2006assessment,nelson2019view}. With the increasingly important role of UAVs in automation and human-adjacent applications, the cost of failure is commensurately higher too. Yet, UAVs are being operated in increasingly stressful and adverse conditions \citep{dhulipalla2022comparative,grishin2020methods,ahmed2023adaptive}. This has increased the need for 
	flight control that is robust to environmental and other disturbances. 
	
	In our review of related work in section \ref{sec:back_uav}, we note that there is a %preponderance of classical and data-driven approaches. However, there is 
	a disconnect between theoretical research with one-off applications, proprietary offerings, and pragmatic open-source robust controllers that are easy to adopt for civilian use. Ebeid, et al. \cite{ebeid2017survey} have reviewed popular open-source flight control software and determined that the preeminent actors in the market are Ardupilot and PX4. Both software applications include embedded low-level controllers running onboard the vehicle. However, as we demonstrate later in this paper, such controllers are not sufficiently robust to environmental disturbances, such as crosswind speeds with drag forces that exceed $5N$. 
	Our solution to this problem is to introduce data-driven reinforcement learning-based supervisory control to make the overall control robust to environmental disturbances, such as crosswinds. However, the computational complexity of deep RL networks grows quadratically as the number of parameters in the deep learning network and the number of data points required to learn the RL-based supervisory controller \cite{bianco2018benchmark}.  The flight computer for UAVs may not have the ability to consistently run control inference at the desired frequency due to limited processing and memory capacity.
	%Both deploy flight control logic on embedded computers with a computational budget that is low for intensive machine learning.
	Thus, any data-intensive machine learning approach must account for the constraints of the infrastructure it is running on.
	
	In this paper, we propose a supervisory reinforcement learning (RL) approach, which develops a path modification framework that is designed to operate in conjunction with the standard autopilot control software to accommodate wind disturbances that may occur during flight. %The approach is interoperable with Ardupilot. 
	The rest of this paper is structured as follows. Section \ref{sec:related} discusses related work in the field of UAV and adaptive control. The reinforcement learning problem is introduced in Section \ref{sec:prelims}. Section \ref{sec:problem} documents features of the UAV control logic, and our problem formulation for adaptive supervisory RL control. Section \ref{sec:uav_nominal} describes the proposed supervisory control solution, along with the Ardupilot control logic that we have not encountered in peer-reviewed literature to date. Finally, we study the importance of hyperparameter optimization in developing the proposed supervisory RL approach and empirically demonstrate its effectiveness for known and unknown adverse wind conditions.  %robust control under adverse conditions.
	
	% =====================================================================
	
	\section{Related work}\label{sec:related}
	\label{sec:back_uav}

	%Small 
	UAVs are  dynamic systems that operate on small time constants ($0.01-0.1$ Hz) and are susceptible to internal and external disturbances during flight. For example, moderate environmental changes and system faults (e.g., loss of a motor) can affect system behavior and destabilize UAV flight. Therefore, it is imperative to develop controllers for UAVs that can adapt to disturbances and faults. Aastrom and Wittenmark \cite{aastrom2013adaptive} define an adaptive controller as one ``with adjustable parameters'' that adapts to dynamic changes in a process. An approach to adaptive control uses a  nested loop: the \textit{inner loop} actuates the system, and the \textit{outer loop} manipulates the operating parameters of the inner loop. Methods such as adaptive model-predictive control (MPC), periodically estimate the process parameters \citep{adetola2009adaptive}. Other approaches, like linear quadratic regulators (LQR) solve a sequence of quadratic  equations with  a single positive-definite solution for controllable systems, to obtain an optimal action based on the linearization of the system at different operating points \citep{bemporad2002explicit}. Prior work on adaptive control in UAVs adopt classical and data-driven approaches.
	
	% -------------------------------------
	\subsection{Classical adaptive control}
	% -------------------------------------
	
	Adaptive PID control \cite{anderson1988rule,tjokro1985adaptive} tunes the controller gains based on measured variables that characterize the system behavior. This makes it easier for the controller to track reference signals. However, this comes at the cost of adaptivity because the disturbances that occur in the system must be known and accounted for ahead of time. Work has been done to make control adaptive by tuning PID gains when a fault is detected \citep{sadeghzadeh2012active}. Similarly, Pounds, et al \citep{pounds2012stability} analyzed the effects of abrupt changes in a UAV's payload and the bounds within which the PID control can adapt. In case of multiple actuator failures, cascaded PID controllers on an octo-rotor can be used to reallocate control to redundant rotors \citep{marks2012control}. There is additional literature on quadrotor autonomy using linear and non-linear controllers %. However several insights are similar across vehicle platforms
	\citep{zulu2014review}.
	
	MPC uses a dynamic model of the system to detect faults and disturbances and provide adaptive control \citep{yu2015mpc}. Residual-based fault detection has been used for fault-tolerant control of a co-axial octocopter \cite{saied2015fault}. The built-in co-axial motor redundancy can be exploited to accommodate faults by increasing motor speeds. More recent research has involved adaptive MPC to estimate process model parameters under uncertainty for fast dynamics \citep{adetola2009adaptive,pereida2018adaptive,chowdhary2013concurrent}.
	
	Other researchers have demonstrated the utility of sliding mode control for fault tolerance in  rotorcraft. Razmi and Afshinfar use sliding mode quad-rotor attitude control with a neural network to learn the sliding mode parameters is demonstrated \cite{razmi2019neural}. Hamadi, et al \cite{hamadi2020comparative} use a self-tuning sliding mode controller to achieve fault tolerance after fault detection.
	
	Some recent work incorporates adaptation built into popular open-source flight control software for fixed-wing vehicles. Baldi, et al \cite{baldi2022ardupilot} use model-free adaptive control for fixed-wind UAV being flown with Ardupilot in wind conditions. They linearize the system dynamics and use a single-step look-ahead cost to adapt the PID controller gains. Similarly, sliding mode adaptive PID gain control has been incorporated for Ardupilot logic for a fixed-wing UAV \cite{li2022embedding}.
	
	% ------------------------------------
	\subsection{Adaptive machine learning-based control}
	% ------------------------------------
	
	Generally, fault-tolerant control schemes using classic control rely on fault detection and isolation as an initial step to adapting the control function to accommodate fault(s) \citep{chamseddine2015active,park2013fault}. Recently, data-driven and RL approaches have been developed for fault-tolerant control leveraging their inherent adaptive capability while attempting to mitigate their sample inefficiency. Usually, they have been employed in conjunction with classic control schemes.
	
	Adaptive classic control literature parallels meta-reinforcement learning  in the field of machine learning \citep{schweighofer2003meta,santoro2016meta}. The inner loop consists of a  \textit{base learner}, which optimizes a single optimal control problem. The outer loop consists of a \textit{meta learner}, which updates the parameters of the inner-loop learner so that it generalizes across varying tasks.
	Recent work by Richards, et al \cite{richards2021adaptive} uses \textit{offline} meta-learning to pre-train an adaptive controller using data-driven simulations. The controller can actuate a drone to follow trajectories under wind disturbances. Razmi and Afshinfar
	\cite{razmi2019neural} use a neural network to tune the parameters of a sliding mode controller for a UAV. The neural network helps address the drawbacks of sliding mode control, such as susceptibility to chattering and measurement noise.
	
	RL approaches are inherently adaptive, as they learn to control by interacting with the  environment. However, the exploration period to allow adaptation may be unsafe and cause failure states. A solution is to use inverse dynamics with General Policy Iteration to learn value functions and then take greedy actions \cite{li2020unmanned}. The sample inefficiency and stochastic exploration of RL are mitigated by constraining the state space to follow only certain trajectories during exploration, which are generated from knowledge of system dynamics.
	
	Similarly, RL controllers have been developed that are robust to actuator and sensor faults (cyber attacks) \cite{fei2020learn}. The RL controller runs concurrently with a cascaded PID controller and adds control signals to the PID controller's outputs. The RL controller was trained in simulation on faults and compensated for sub-optimal PID actions when faults occurred. This approach was robust and did not need separate fault detection and isolation. 
	
	On a similar note, control logic may be abstracted away from systemic changes where possible, to make control adaptation faster \cite{pi2021general}. A system-agnostic  UAV control scheme for quad- and hexacopter platforms using RL has been developed. The controller outputs the needed translational and rotational accelerations, which are then allocated to the rotors via a control allocation method that corresponds to the individual vehicle's geometry. This approach demonstrated the utility of data-driven approaches in abstracting knowledge between tasks for fast adaptation when needed.
	
	Another approach combines neural networks with a physics-based model \citep{shi2019neural}. The physics-based model learns the lower-order dynamics, whereas the neural network models higher-order effects like noise and the effects of faults on the nominal dynamics. This provides the basis for more accurate model-based control.
	
	In summary, data-driven approaches have been used to complement classic control techniques. The ability to learn from experience and the expressive power of neural networks to represent complex control laws allows for satisfactory generalization. However, in many cases, machine learning approaches to adaptive control may not be practical in real-world applications due to computational and execution time constraints as well as mismatched input/outputs between the proposed solution and extant flight control software. In the work that follows, we develop an RL control scheme that is robust to wind disturbances and can operate in a supervisory mode in conjunction with
	%popular hobbyist and 
	research flight software.
	
	% =====================================================================
	
	\section{Reinforcement learning (RL)}\label{sec:prelims}
	
	This work discusses the RL-based control of systems modeled as Markov Decision Processes (MDPs). We define an MDP control task characterized by the process dynamics $P: X \times U \rightarrow X$, which maps the current state of a system, $x_i$ to its next state $x_{i+1}$, given an input $u_{i+1}$. MDPs adhere to the Markov Property. That is, given a state and an action, the state transition probability is conditionally independent of prior states and actions. Or, each state encodes sufficient information to predict the next state, given an action. Each state transition incurs a reward that is based on the state and the action taken to reach that state $r: X \times U \rightarrow \mathbb{R}$. An episode is a sequence of interactions (i.e., inputs) that takes the system from a current state to a terminal goal state. The objective for the control task is to derive a policy $\pi: X \rightarrow U$, such that each action picked maximizes the expected discounted future returns $\mathbb{E}[G(x_i)]$, where $G(x_i) = \sum_{j=i}\gamma^{j-i}r(x_i,\pi (x_i))$. The discount factor, $\gamma \in (0,1]$ prioritizes the immediacy of feedback. Expected discounted returns in a given state $x$ under an optimal policy are known as its value, $V(x)=\max_{u}\mathbb{E}[G(x)]$. In these situations, the control problem can be framed as $\pi_t(x) \gets \arg \max_u V(x)$.
	
	% =====================================================================
	
	\section{Supervisory control of UAVs}\label{sec:problem}
	
	This section discusses the specific research problem addressed in this work. First, we discuss the control logic of a UAV implemented in the form of the popular Ardupilot software. Ardupilot uses a cascaded PID architecture for converting waypoint references to control signals for the motors. Secondly, the broader problem of navigation is described in real-world conditions.
	
	% -----------------------------------
	\subsection{UAV control with PID}
	% -----------------------------------
	
	\begin{figure}
		\centering
		\includegraphics[width=0.5\textwidth]{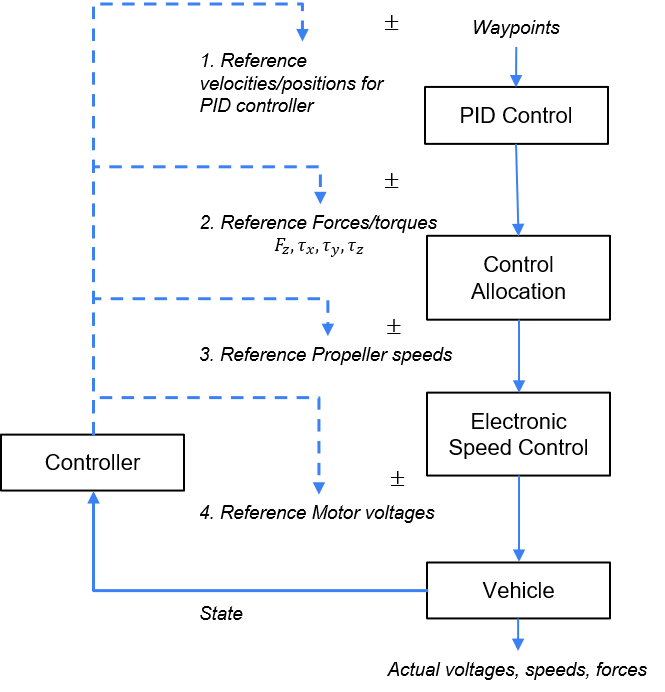}
		\caption{The control logic flow for a UAV.}
		\label{fig:uav_overview}
	\end{figure}
	
	The standard control logic flow of UAVs is depicted in Figure \ref{fig:uav_overview}. In this standard end-to-end approach, the position reference is converted into propeller speed signals.
	
\begin{figure}
		\centering
		\includegraphics[width=0.8\textwidth]{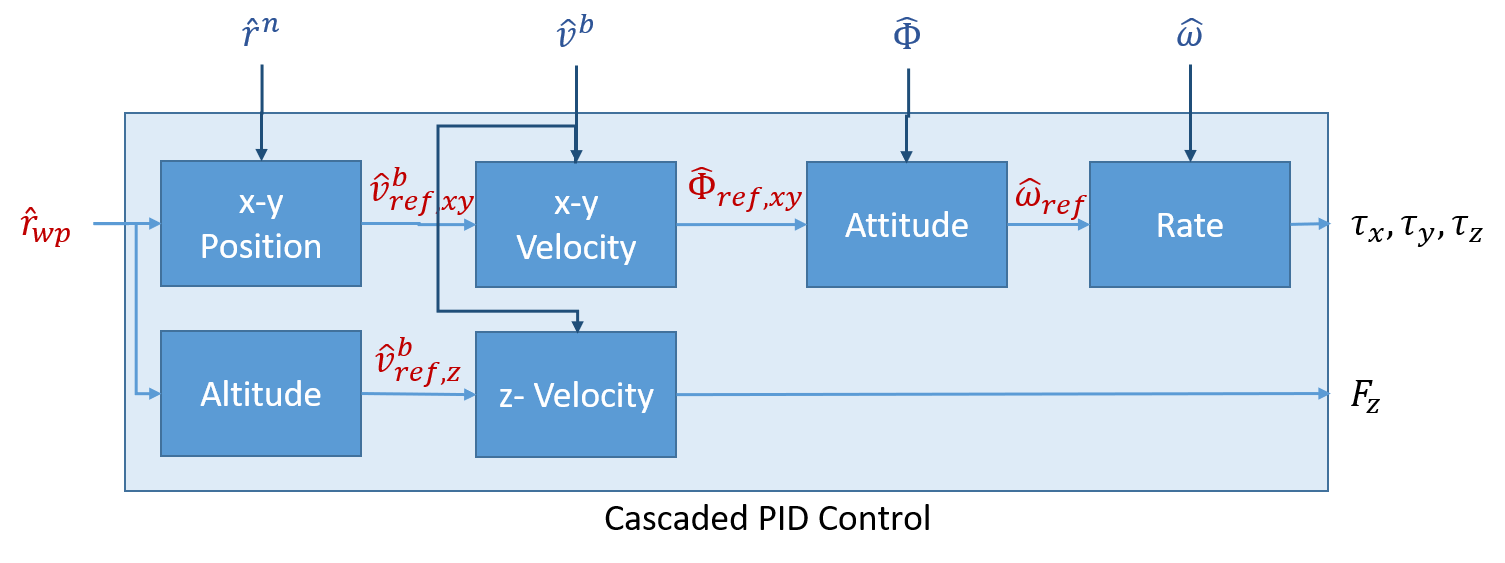}
		\caption{The structure of a cascaded PID controller. Position error is eventually converted to desired thrust and torques.}
		\label{fig:cascaded_pid}
	\end{figure}

	The first step is PID control (see Figure \ref{fig:cascaded_pid}) converts position references into the desired dynamics (forces and torques, $F_z,\tau_x,\tau_y,\tau_z$) to achieve that reference position. The underlying control for position tracking follows a cascaded PID architecture. The control output $u$ aims to minimize the input error $e=\texttt{reference}-\texttt{measurement}$ with respect to the reference and measured states:
	
	\begin{align}\label{eqn:pid}
	u_{PID} = k_p e + k_d \nabla_t e + k_i \int_0^t e \partial t
	\end{align}
	
	This PID setup simultaneously tracks lateral and vertical motion. For lateral motion along the $x,y$ axes, the position error is converted into a velocity reference. The velocity controller outputs the required roll and pitch angles, which are then converted to the corresponding angular rate. Finally, the rate controller converts the prescribed angular rate into the torques needed from the motors. The cascaded PID controllers operate at different frequencies. For our experiments, the attitude and rate controllers operate at $100$Hz to keep up with higher-level position and velocity controllers that operate at a slower rate of $10$Hz. In other words, the attitude control is able to catch up with its reference and reduce error before the reference is changed by position control. This is typical of the inner loop of cascaded control architectures.
	
	However, the canonical form of the PID controller in equation \ref{eqn:pid} is susceptible to integral wind-up, noisy inputs, and large actions that are unsafe. The Ardupilot project for autopilot navigation has developed mitigation logic to address these conditions for lateral position control.
	
	\textit{Square-root scaling} the proportional error term $k_p$ for lateral position and attitude control avoids overshoot in reacting to large initial errors when approaching the reference point. Square-root scaling is a function of  the maximum allowed acceleration of $e$, i.e., $\Ddot{e}=\partial^2 e/ \partial t^2$ and the time interval of control action $\Delta t$. Equation \ref{eqn:sqrt_scaling} shows that if the error exceeds a threshold proportional to $\Ddot{e}/k_p^2$, the proportional response is scaled with respect to the square root of $e$. % exceeding said threshold.
	
	\begin{align}\label{eqn:sqrt_scaling}
	u = \texttt{clip} \left(\begin{cases}
	k_p=0   & 
	\begin{cases}
	e>0 &   \sqrt{2 \Ddot{e} e} \\
	e<0 & -\sqrt{2 \Ddot{e} (-\Ddot{e})}
	\end{cases} \\
	k_p>0   &
	\begin{cases}
	e>\frac{\Ddot{e}^2}{k_p^2}     &   \sqrt{2 \Ddot{e} (e - \frac{\Ddot{e}^2}{2k_p^2})} \\
	e<-\frac{\Ddot{e}^2}{k_p^2}    &   -\sqrt{2 \Ddot{e} (-2 - \frac{\Ddot{e}^2}{2k_p^2})} \\
	e=0             &   k_p e
	\end{cases}
	\end{cases}, -\frac{|e|}{\Delta t}, \frac{|e|}{\Delta t}
	\right)
	+ k_d \nabla_t e + k_i \int_0^t e \partial t
	\end{align}
	
	When \textit{leashing} is applied to the lateral position controller, it prevents the error input to the PID from becoming too large. As equation ~\ref{eqn:leash} shows, the position error is clipped to a maximum value given the current kinematics of the system.
	
	\begin{align}
	\label{eqn:leash}
	\texttt{leash} = \left| \frac{\Ddot{|e|}}{2 k_p^2} + \frac{\Dot{|e|}^2}{2 \Ddot{|e|}} \right|
	\end{align}
	\noindent
	Thus a leash can be used to threshold the maximum error fed to the square-root scaling function, which in turn modulates the proportional response of the PID controller.
	
	Once the cascaded PID setup outputs desired dynamics, the control allocation step converts them into desired propeller speeds. The speeds are then converted to voltage signals for input to the motors by the electronic speed controllers (ESCs), and this actuates the vehicle.
	
	Such a control approach presents multiple interaction points for interleaving supervisory control using a reinforcement learning controller. For example, as shown in Figure \ref{fig:uav_overview}, a supervisory controller may modify the trajectory waypoints being fed to the cascaded PID controller. Alternately, the prescribed forces and torques may be modified to optimize for some navigation objectives before they are allocated as speeds. Similarly, the voltage signals to the speed controllers may be changed. In summary, theoretical and practical factors limit the choices of where adaptive control schemes may be applied.
	
	\subsection{Trajectory control of UAVs}
	
	Our overall objective in this work is optimal trajectory control of the UAV in real-world operations, where flights may be disrupted by disturbances, such as wind. In this paper, we modify the control architecture to ensure that the originally intended trajectory is followed as closely as possible in spite of these disturbances. 
	
	Reactive controllers, such as PID, can use tailored parameters to adjust to different operating conditions. However, they are designed to follow the original reference trajectory, specified for flights with no environmental disturbances. As a result, a PID position controller in a headwind may fall short of the originally specified reference position unless the proportional and integral terms are amplified. However, such amplifications also cause strong reactions to errors, which can cause overshoot. %However, the same amplified terms will cause overshoot in a different direction. 
	We overcome this problem by designing supervisory controllers that are proactive and can alter the reference trajectory to accommodate the disturbance.
	
	Proactive control methods, such as MPC, can use an accurate model of the environment to replan ahead. However, system identification and the fine-tuning of parameters increase the computational load of the control algorithm. In addition, solving the limited horizon optimization problem for each decision can become intractable for systems with small time constants, because they require very fast reactions.
	
	RL can also be used for proactive control. Developing an RL controller is a two-step process. In the learning step, RL solves an optimization problem to learn the control policy by taking exploratory actions and observing feedback. The control policy, once learned, can generate the best control actions to take for a given state of the system. This, like MPC, addresses the problem of reactivity in PID control. It also ameliorates the recurrent computational costs associated with the MPC optimization problem. The learning process, however, suffers from the challenges of safety because of the stochastic actions used for learning, the large amounts of data that may be needed to learn a good policy, and the computational cost of hyperparameter optimization. Therefore, an RL controller that does not need to learn often under different conditions presents one solution. Such a controller will be robust to disturbances. A robust controller will fulfill the flight objective under different disturbance conditions. In the following section, we present the architecture and the design choices for developing a robust supervisory controller when the UAV system operates under disturbances, such as fixed wind conditions.
	
	% ==============================================================
	
	% ===============================
	\section{Supervisory control solution formulation}\label{sec:uav_nominal}
	% ===============================
	
	We develop a waypoint-to-waypoint supervisory strategy for training a robust controller, $\pi_{RL}$, and using that controller for trajectory control scenarios. Our proposed approach divides the robust controller's development into a training and an operating phase. During the training phase, the controller learns to fly the vehicle from different initial states in a 3D $\texttt{bounding box}$ to the origin. For each controller's training the prevailing wind disturbance conditions are kept constant. During operation, the desired path, comprising position and yaw references, is broken down into intermediate waypoints $\hat{r}_{wp}$. The RL controller's position is set relative to the waypoint, Then the controller navigates to the relative origin, while ultimately flying to $\hat{r}_{wp}$. 
	
	The UAV's control and dynamics processes, described previously are modeled as a MDP that is controlled by RL. The actions to the MDP are the changes in position references being tracked by the PID. The state observation is the kinematic state of the vehicle. The reward being optimized for by RL incentivizes short, straight flight segments to reference positions along the pre-specified trajectory (see Figure~\ref{fig:uav_problem}). Section \ref{sec:uav_action_space}, Section \ref{sec:uav_state_space}, and Section \ref{sec:uav_reward} explain the choice of these parameters in greater detail. Then Section \ref{sec:uav_training} and Section \ref{sec:uav_operation} explain our proposed algorithms for training and operating a robust controller. 
	
\begin{figure}[ht]
		\centering
		\includegraphics[width=0.8\textwidth]{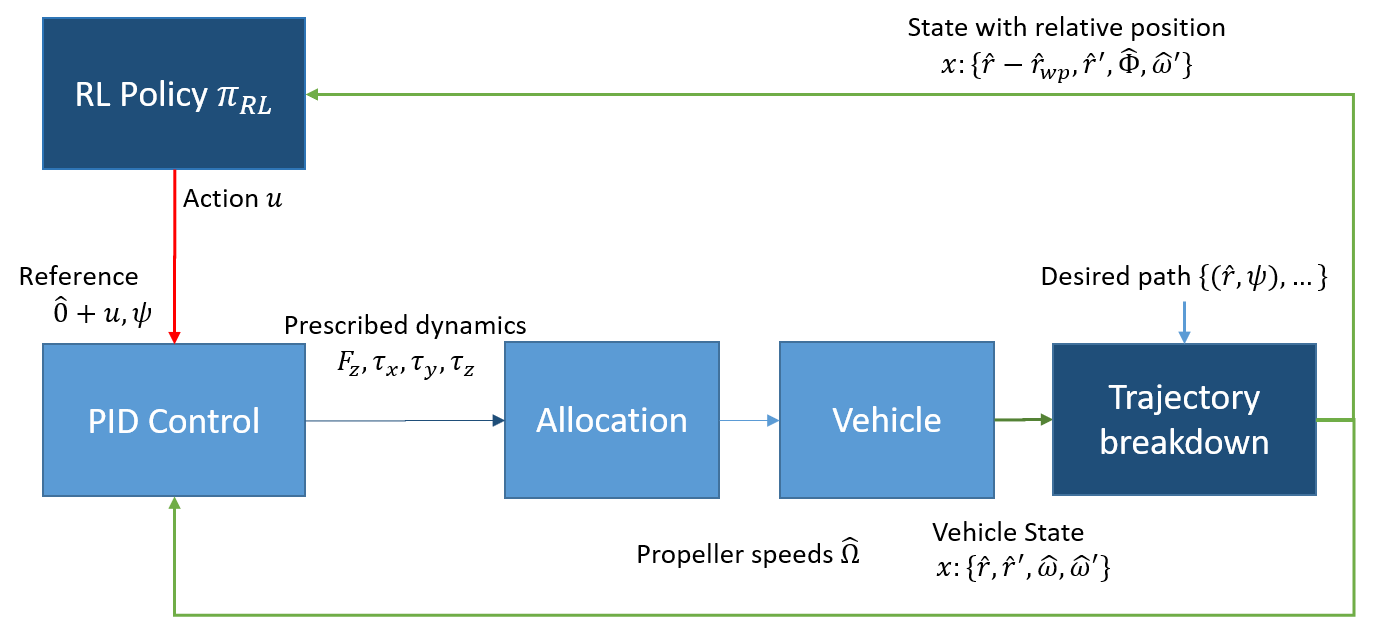}
		\caption{The UAV Trajectory control interface via waypoint control. The reference positions being tracked by the PID controller are modified.}
		\label{fig:uav_rl_traj_control}
	\end{figure}

	Figure \ref{fig:uav_rl_traj_control} shows how the proposed supervisory RL controller interacts with the vehicle's control systems. Our proposed approach introduces two supervisory blocks. Trajectory breakdown decomposes the desired path into waypoints, and changes the position in the navigation frame to a position relative to the next waypoint for the RL controller. The RL controller acts in a supervisory capacity without interrupting the logic of the cascaded PID controller. The changes made to the reference of the PID control cascade through the system. This has the practical advantage of being minimally invasive and easy to implement with existing flight control software like Ardupilot. Secondly, the RL controller is based on a neural network, which is a function approximator for the control policy. It is computationally more demanding and may not be able to run at higher frequencies of the embedded control logic. Having RL in a supervisory capacity, which modifies variables of the low-level control loop at a lower frequency, prevents it from being a performance bottleneck in time-constrained applications, such as this.
	
	In the following subsections, we define the state and action spaces of the control task, and the reward function being optimized. Finally, the overall experimental setup is discussed. The results of our experimental studies are presented in Section \ref{sec:uav_results}.
	
	\subsection{Action space}\label{sec:uav_action_space}
	
	We model supervisory actions as additive modifications to the waypoint of the navigation problem. The controller's objective is to modify the waypoint from the origin, such that the effects of disturbances, which are also additive (especially in the case of simple wind profiles) are negated. The supervisory controller operates at a lower frequency than the underlying PID control.
	
	Theoretically, the choice of control action $u$ and the frequency of application of control $f_{RL}$ is a function of the computational tractability of the corresponding control optimization problem. Given that the internal PID controller of the UAV has its own transient state, the UAV system dynamics do not satisfy the Markov property (section \ref{sec:prelims}). This is because the defined UAV system state is defined only by the kinematic variables of the vehicle's motion  (see section \ref{sec:uav_state_space}). In other words, the system is partially observable. There are two approaches to mitigating partial observability of the state space. In the first approach, we append the variables of the PID controller to the state vector of the vehicle, which is input to the RL agent. In the second approach, we lower the frequency $f_{RL}$ of the interaction of the RL agent with the vehicle. The lower frequency of interactions results in the transient effects of changing the references inside the PID controller to die down. 
	
	In this paper, we adopt the second approach. The RL controller takes a supervisory action after $(f_{RL} \delta t)^{-1}$ steps, where $\delta t$ is the simulation step computed from the time constant of the vehicle dynamics. It allows the RL agent to view the state of the vehicle after the PID controller has actuated the vehicle to follow the reference position. In this way, the RL agent can train on state observations that are correlated with its control outputs.
	
	Practically, the choice of control action is constrained by the problem application. We envision combining the RL-based supervisory controller with Ardupilot-based control operating in the Guided Mode. The Ardupilot-based controller operates on the flight computer. In the Guided Mode, the low-level controller accepts position and velocity references from the ground station and outputs the motor control signals to the electronic speed controllers. Due to the lag associated with communication, and the possibility of connection loss, actions are taken in a supervisory manner wherein a baseline level of control is possible without RL intervention. Having such a supervisory controller makes interoperability with Ardupilot easier, without having to interrupt its internal control logic.
	
	\subsection{State space}\label{sec:uav_state_space}
	
	The UAV is modeled with dynamic state variables, which describe its flight \citep{ahmed2022high}. These variables, listed below, describe the kinematic state of the vehicle:
	
	\begin{itemize}
		\item $\hat{r}^n \in \mathbb{R}^3$ are the navigation coordinates, in the global, inertial navigation reference frame axes $n$.
		\item $\hat{v^b} \in \mathbb{R}^3$ is the velocity of the vehicle in the vehicle's local, non-inertial body frame $b$.
		\item $\hat{\Phi}=[\phi, \theta, \psi] \in [-\pi, \pi]^3$ is the orientation of the body reference frame $b$ in Euler angles (roll, pitch, yaw) with reference to the inertial reference frame $n$.
		\item $\hat{\omega}=[\omega_x, \omega_y, \omega_z] \in \mathbb{R}^3$ is the roll rate of the three body frame axes with reference to the inertial frame.
	\end{itemize}
	\noindent
	We assume they also represent the state space for the controller.
	
	For the reasons discussed above, i.e., the PID controller contributing to the state describing the dynamics of the UAV, an action may produce different dynamics. %For example, if the rate of change of errors $\hat{e}$ are different. 
	Thus, to satisfy the Markov property, such that the RL controller can evaluate optimal actions, the internal state of the PID should be a part of the state vector input to the controller. However, there are 6 cascaded PID controllers, each with an internal derivative and integral state. This amounts to 12 additional variables, doubling the dimensionality of state space. This makes the search space for an optimal policy exponentially larger. 
	
	Furthermore, as explained in our choice of actions, the state of the PID is transient. At constant velocity and infrequent supervisory actions, the rate of change of error will be constant. And clipping the integral error to low magnitudes will prevent the effects of wind-up. Considering this transient nature, and infrequent actions, limiting the state vector to the kinematic variables presents a compromise between accuracy and computational tractability.
	
	\subsection{Reward}\label{sec:uav_reward}
	
	The objective (or, reward) is to reach a waypoint $\hat{r}_{wp} \in \mathbb{R}^3$ in the shortest possible time from the starting position $\hat{r}_{0}$ without violating velocity constraints. The implication of this is to traverse the shortest distance and at the greatest allowable speed. Therefore, the reward function imposes a constant time penalty for each interaction with the environment. Given an objective waypoint $\hat{r}_{wp}$, the last and current positions $\hat{r}_0,\hat{r}_1$, it rewards motion towards the waypoint and penalizes motion perpendicular to the shortest path. The reward formulation is further elaborated in equation \ref{eqn:uav_reward}.
	
	\begin{align}
	r(\hat{r}_i, \hat{r}_{i+1}, \psi_i, \psi_{i+1}) = 
	-\delta t - \left(|\psi_{i+1}| - |\psi_{i}|\right)
	+ |(\hat{r}_{i+1} - \hat{r}_i)|
	- |(\hat{r}_{i+1} - \hat{r}_i) \times (\hat{r}_{wp} - r_0)|
	\label{eqn:uav_reward}
	\end{align}
	
	This reward formulation is made up of several components. The time penalty $\delta t$ is a constant penalty for each additional time step before reaching the destination (i.e., the next waypoint). A turn penalty $|\psi_{i+1}| - |\psi_{i}|$ incentivizes the controller to keep a constant heading by penalizing yaw. For this experiment, position control is only done via roll and pitch changes ($\phi, \theta$). An ``advance" term $|(\hat{r}_{i+1} - \hat{r}_i)|$ rewards motion, towards or away from the target. In our experiments for finding a suitable reward function, we noted that a \textit{signed} advance term yields worse results. This may be because a signed advance term confounds objectives with the time penalty. Advancing \textit{towards} the target necessarily will accumulate a smaller time penalty. However, sometimes, due to an initial velocity away from the destination, the controller will necessarily accrue negative advance as it decelerates. Finally, a ``cross" deviation term $|(\hat{r}_{i+1} - \hat{r}_i) \times (\hat{r}_{wp} - r_0)|$ penalizes motion perpendicular to the position vector, the shortest path, to the destination. The ending condition of an episode incurs an additional bonus reward or penalty (equation \ref{eqn:terminal_reward}).
	
	\begin{align}\label{eqn:terminal_reward}
	r \leftarrow r + 
	\begin{cases}
	\texttt{reached}    &   + \texttt{bounding box} \times 20 \\
	\texttt{tipped} (\Phi > 30\degree) \texttt{ or out-of-bounds}    & - \texttt{bounding box} \times 20 \\
	\texttt{timed out (t > 20s)}    & -|\hat{r}_{i+1} - \hat{r}_{wp}| \times 10 
	\end{cases}
	\end{align}
	
	\subsection{Experiment setup}
	
	We use a simulation testbed for a Tarot T-18 octo-rotor UAV. For this simulation environment, the maximum roll and pitch angles ($\phi,\theta$) are restricted to $\pi/12$ radians. Lateral motion is achieved via roll and pitch. Therefore, yaw controls are disabled. The maximum velocity is restricted to $3 ms^{-1}$, and the bounding box for the granular navigation problem is $20m$ on all sides. The bounding box dimensions are discussed in greater detail in subsequent subsections. The maximum lateral acceleration the vehicle can achieve is 2.5 $ms^{-2}$, given the maximum tilt angle and a maximum propeller speed, leading to a net lateral force of $26.65 N$, given the UAV mass is $10.66 kg$.
	
	Disturbances are modeled as a wind force with a constant heading applied through the duration of a trajectory. We assume the effects of turbulent wind are minimal, and the duration of the flight is within the time frame of the constant prevailing wind. We conduct experiments with a maximum lateral wind disturbance of $5N$, accounting for $18.7\%$ of the available lateral thrust.
	
	Furthermore, a set of PID-only controller parameters is first obtained for adequate comparison. The parameter space is tabulated in Tables \ref{tbl:pid_params}, \ref{tbl:rlpid_params}. The relationships between parameters and the performance as measured by robustness are discussed for each set of experiments. Other experimental details are presented in Appendix \ref{sec:appendix}.
	
	An episode represents a single, granular, tracking problem. The vehicle is initialized at a position and velocity relative to the origin. At each step, the controller receives the reward for its state. The episode ends when the vehicle comes within a propeller arm span of the origin, times out, exceeds the position bounding box, or tips over. A larger trajectory is simulated by solving a sequence of navigation problems and adding up relative positions.
	
	The Proximal Policy Optimization (PPO) algorithm is used for RL \cite{schulman2017proximal}. PPO operates over the continuous state and action spaces. It clips the magnitude of iterative updates to avoid drastic changes in the policy. For learning a policy, the state vector is min-max normalized to the range $[-1,1]$. This is to prevent an asymmetric cost surface for RL optimization. That may cause activation function saturation from state variables that become too large, or gradient updates to policy parameters that overshoot the optimum \cite{sola1997importance}.
	
	\subsection{Training phase}\label{sec:uav_training}
	
\begin{figure}[ht]
		\centering
		\includegraphics[width=0.8\textwidth]{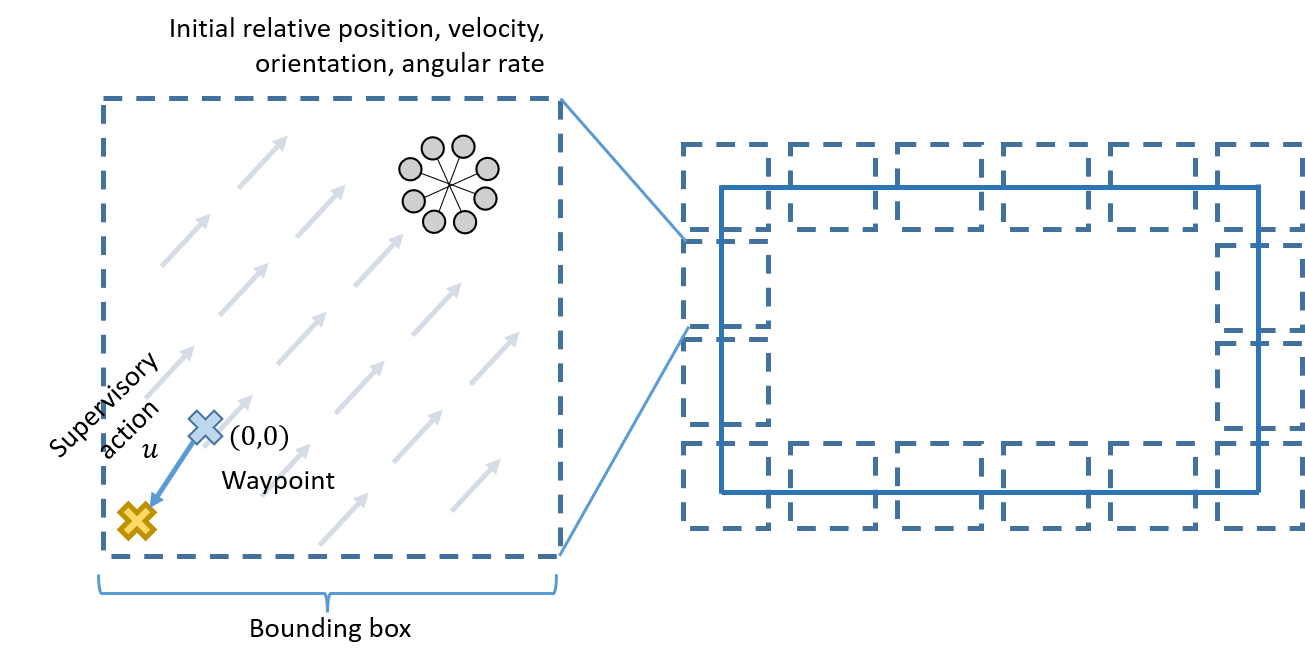}
		\caption{The trajectory modification problem under disturbances such as wind. The goal is to break up a longer trajectory into smaller, granular problems (left), where the supervisory controller changes the relative waypoint to counteract disturbances.}
		\label{fig:uav_problem}
	\end{figure}

	The controller is trained by presenting it with ``episodes" of granular navigation problems. For each problem, the vehicle is initialized with a starting position and velocity inside a 3D $\texttt{bounding box}$, along with a constant wind disturbance force. For each navigation problem, the relative waypoint is the origin. Since the eventual goal is trajectory control over longer paths, Figure \ref{fig:uav_problem} illustrates how a longer trajectory may be broken up into a sequence of such episodes. This is done by initializing every subsequent episode with the terminal state of the vehicle from the prior episode and changing its position relative to the next waypoint. We hypothesize that by solving a series of smaller navigation problems, we are able to reduce the problem complexity and also counteract the effects of disturbances over longer trajectories.
	
	The supervisor modifies the reference position being tracked by the cascaded PID controller, which continually actuates the vehicle. The amount by which the waypoint reference can be modified is limited to the size of the $\texttt{bounding box}$ of the episode during training. The $\texttt{scaling factor} \in (0,1]$ is a hyperparameter for the controller, which further limits the amount by which the reference waypoint can be modified inside that bounding box. The actual change in reference waypoint is $ u \leftarrow \texttt{scaling factor} \cdot \texttt{bounding box} \cdot u$.
	
	The choice of the size of the $\texttt{bounding box}$ is a design-time decision for this work. A smaller bounding box, given the flight envelope of the vehicle, namely maximum acceleration and velocity, may not provide sufficient room to navigate against disturbances, for example when the vehicle is blown off-course by the wind. A larger bounding box will make the state space (and policy search space) larger with little marginal benefit. For a large bounding box, a bulk of the flight in a bounding box will be spent at constant velocity from saturated PID control outputs due to a large position error. Therefore, the relationship between supervisory changes to position references and the vehicle's response will have no correlation. Making the bounding box smaller thus increases the efficiency of the RL policy search process.
	
	Algorithm \ref{alg:uav_rl_training} illustrates the steps for training a supervisory controller using reinforcement learning.

	\begin{algorithm}[htb]
		\caption{Training supervisory RL control}
		\label{alg:uav_rl_training}
		\begin{algorithmic}[1] %[1] enables line numbers
			\Require Wind disturbance force $F_{xy} \in \mathbb{R}^2$
			\Require UAV model with state $x$: position $\hat{r}$, velocity $\hat{v^b}$, orientation $\hat{\Phi}$, angular rate $\hat{\omega}$
			\Require number of training episodes, $\texttt{bounding box, scaling factor}, f_{RL}$
			\State Initialize supervisory controller $\pi_{RL}: x\rightarrow [-1,1]^3$
			\For{episode in episodes}
			\State Set random UAV position $\hat{r}$ inside $\texttt{bounding box}\in\mathbb{R}^3$
			\State Set random UAV velocity $\hat{v^b}$
			\While{$\hat{r} \neq \hat{0}$}
			\If{time more than $1/f_{RL}$ since last action}
			\State Set position reference $u \gets \hat{0} + \texttt{bounding box} \cdot \texttt{scaling factor} \cdot \pi_{RL}(x)$
			\EndIf
			\State Get prescribed dynamics from PID, given reference
			\State Allocate propeller speeds
			\State Apply $F_{xy}$ to UAV model
			\State Update state measurement of UAV $\{\hat{r},\hat{v^b},\hat{\Phi},\hat{\omega}\}$
			\State Observe reward $r$
			\EndWhile
			\State Collect $x,u,r$ experiences and update $\pi_{RL}$ using PPO algorithm
			\EndFor
			\State Return $\pi_{RL}$
		\end{algorithmic}
	\end{algorithm}
	
	\subsection{Operating phase} \label{sec:uav_operation}
	
	During operation, a longer desired path is broken up into intermediate waypoints $\hat{r}_{wp}$, which are a $\texttt{bounding box}$ apart. For each step of operation, the state of the vehicle observed by the RL controller is changed, such that the position is relative to the waypoint. Thus, the relative destination is the origin in the controller's frame. This then becomes the same navigation task as during training. The controller sequentially navigates through the waypoints, until it finishes the desired trajectory.Algorithm \ref{alg:uav_rl_operation} delineates the proposed approach for the trajectory control problem.
	
	The $\texttt{bounding box}$ parameter also has ramifications for safety constraints of operation, in addition to the efficiency of machine learning. By harshly penalizing deviations away from the waypoint that exceed that box, the controller is given a soft-constraint for adhering to a safety corridor in flight.
	
	\begin{algorithm}[htb]
		\caption{Supervisory RL control for trajectory following}
		\label{alg:uav_rl_operation}
		\begin{algorithmic}[1] %[1] enables line numbers
			\Require Trajectory of waypoints, $\texttt{bounding box}, \texttt{scaling factor}$
			\Require UAV with state $x$: position $\hat{r}$, velocity $\hat{v^b}$, orientation $\hat{\Phi}$, angular rate $\hat{\omega}$
			\Require Supervisory controller $\pi_{RL}: x\rightarrow [-1,1]^3$
			\State Break trajectory into waypoints $\texttt{bounding box}$ apart
			\For{$\hat{r}_{wp}$ in waypoints}
			\State Set relative position of UAV to $\hat{r}_{wp} - \hat{r}$
			\While{$\hat{r} \neq \hat{0}$}
			\If{time more than $1/f_{RL}$ since last action}
			\State Set position reference $\gets \hat{0} + \texttt{bounding box} \cdot \texttt{scaling factor} \cdot \pi_{RL}(x)$
			\EndIf
			\State Get prescribed dynamics from PID, given reference
			\State Allocate propeller speeds
			\State Update state measurement of UAV $\{\hat{r},\hat{v^b},\hat{\Phi},\hat{\omega}\}$
			\EndWhile
			\EndFor
		\end{algorithmic}
	\end{algorithm}
	
	% ==============================================================
	
	% ===============================
	\section{Results and Discussion}\label{sec:uav_results}
	% ===============================
	
	In this section, the performance of our supervisory RL control approach is evaluated in terms of robustness to wind disturbances. A hyperparameter search is conducted to find the best-performing parameters for each experiment. The choice of hyperparameters is evaluated by averaging the total reward over 10 episodes. First, we evaluate the baseline case of PID and RL control under nominal conditions. Then, optimal hyperparameters for PID control under disturbances are evaluated. Finally, an RL supervisory controller is learned for different disturbance conditions. We demonstrate how the controller generalizes to novel conditions without substantial degradation in performance.
	
	% -----------------------------------------
	\subsection{Hyperparameter optimization for nominal control}
	% -----------------------------------------
	
\begin{figure}[ht]
		\centering
		\begin{subfigure}{\textwidth}
			\centering
			\includegraphics[width=1.0\textwidth]{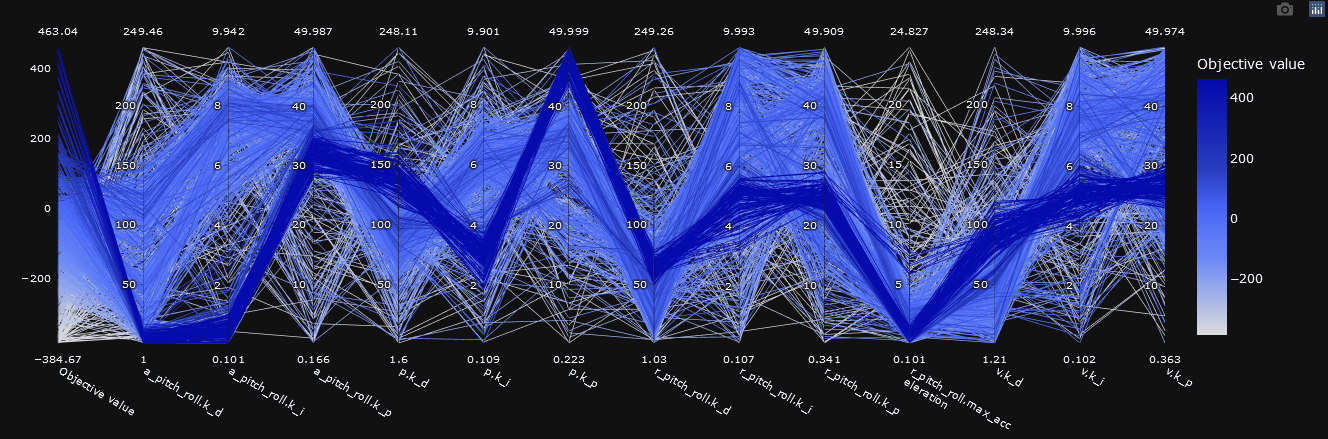}
			\caption{Cascaded PID only.}
			\label{fig:uav_pid_parallel}
		\end{subfigure}
		\begin{subfigure}{\textwidth}
			\centering
			\includegraphics[width=1.0\textwidth]{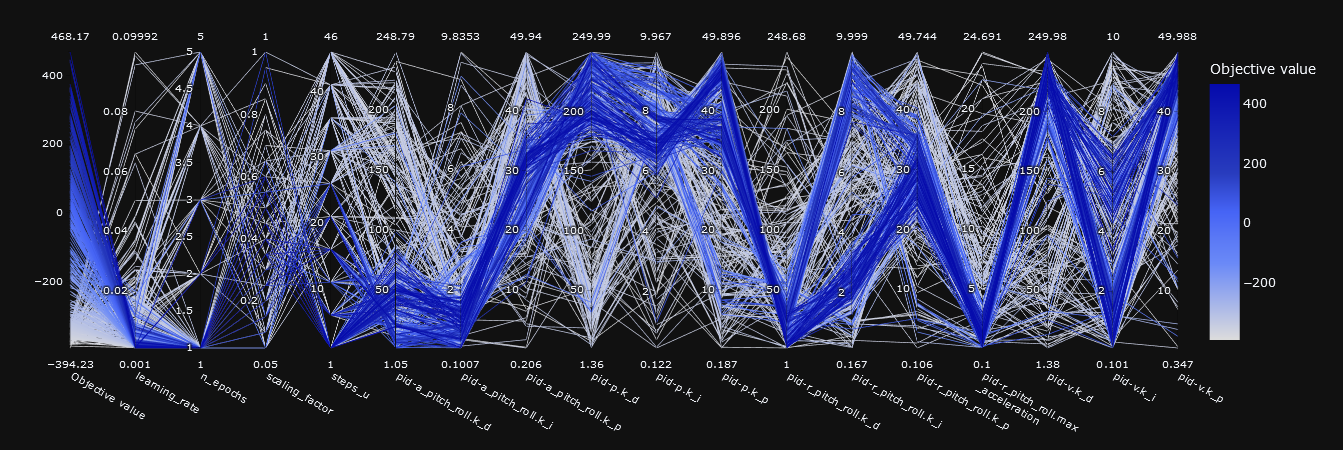}
			\caption{Supervisory RL and cascaded PID.}
			\label{fig:uav_rlpid_parallel}
		\end{subfigure}
		\caption{A parallel coordinate plot of controller parameter combinations, shaded by mean total reward over the same 10 episodes in nominal conditions. The darker tract of lines indicates the combinations that performed well on the navigation task.}
	\end{figure}

	Figure \ref{fig:uav_pid_parallel} shows the result of 1000 hyperparameter optimization trials. For each trial, the search space is over $k_p, k_i, k_d$ parameters of the cascaded PID controller. Another hyperparameter is the maximum acceleration allowed in the rate controller, $max_{acc}$. For the position, velocity, and rate controllers, the most important \citep{pmlr-v32-hutter14} parameter is $k_p$. For the attitude controller, however, the $k_d$ parameter has a higher importance as the controller damps the proportional response so the vehicle does not tip over the reference attitude angle. For the rate controller, and overall, the most important parameter is the maximum allowed angular acceleration. It governs the maximum prescribed torque output to the control allocation block. A higher maximum acceleration will cause the propellers to suddenly generate a larger torque, causing the vehicle to tip over.
	
	In our experiments for hyperparameter optimization, we noted that re-using optimal PID parameters from the PID-only control case for the RL control case often led to sub-optimal results. By changing the position references marginally, the RL controller  modifies the position error, as well as the downstream velocity and angular rate errors to achieve the optimization objective. Therefore,  a PID controller with error constants tailored to accommodate the scale of such reference changes is needed. As a result, the hyperparameter search for the RL supervisory case also searched over the space of underlying PID parameters.
	
	Figure \ref{fig:uav_rlpid_parallel} shows the results of 1000 hyperparameter optimization trials under no wind disturbances for RL control. Like the PID-only case, the most important parameter remains the maximum acceleration limit of the rate controller. $k_p$ is the most important parameter for the position controller. $k_d$, for the velocity, attitude, and rate controllers. We hypothesize that $k_d$ is more important in the supervised case to damp the changes in the tracking error due to the RL supervisor intermittently changing waypoints. 
	
\begin{figure}[ht]
		\centering
		\begin{subfigure}{\textwidth}
			\includegraphics[width=\textwidth]{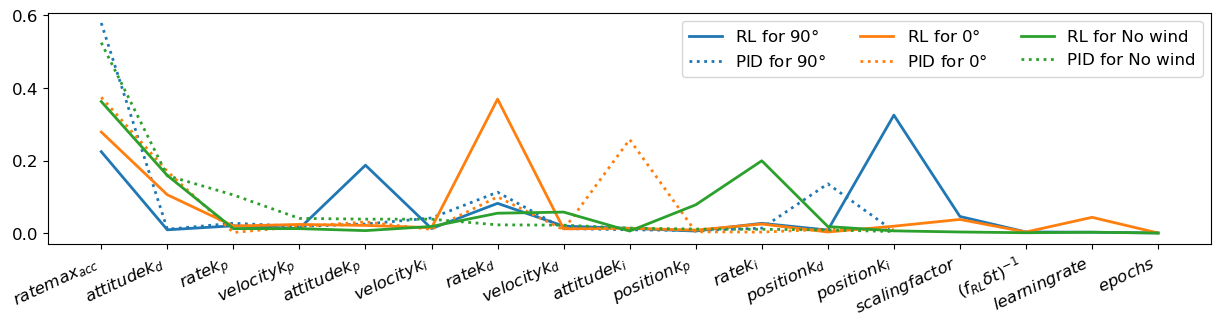}
		\end{subfigure}
		\caption{hyperparameter importance \citep{pmlr-v32-hutter14} for RL and PID control with various wind disturbances. The x-axis is ordered in descending order of parameter importance for the PID-only case under no wind.}
		\label{fig:uav_param_importance}
	\end{figure}

	Figure \ref{fig:uav_param_importance} shows the relative parameter importance for the hyperparameter searches across all RL and PID controllers. The most important parameter overall with respect to the scoring metric is the maximum angular acceleration (rate $\max_{acc}$), which limits the maximum action output of the Rate controller $(\tau_x,\tau_y, \tau_z)$. A large torque as a result of a sudden change in reference can cause the vehicle to tip over, which incurs a large penalty.
	
	% ---------------------------------
	\subsection{PID control with wind}
	% ---------------------------------
	
	Next, the derived controllers are tested with wind disturbances. For the duration of the flight, the wind is considered blowing from a constant heading and with a constant magnitude.
	
\begin{figure}[ht]
		\centering
		\begin{subfigure}{\textwidth}
			\centering
			\includegraphics[width=1.0\textwidth]{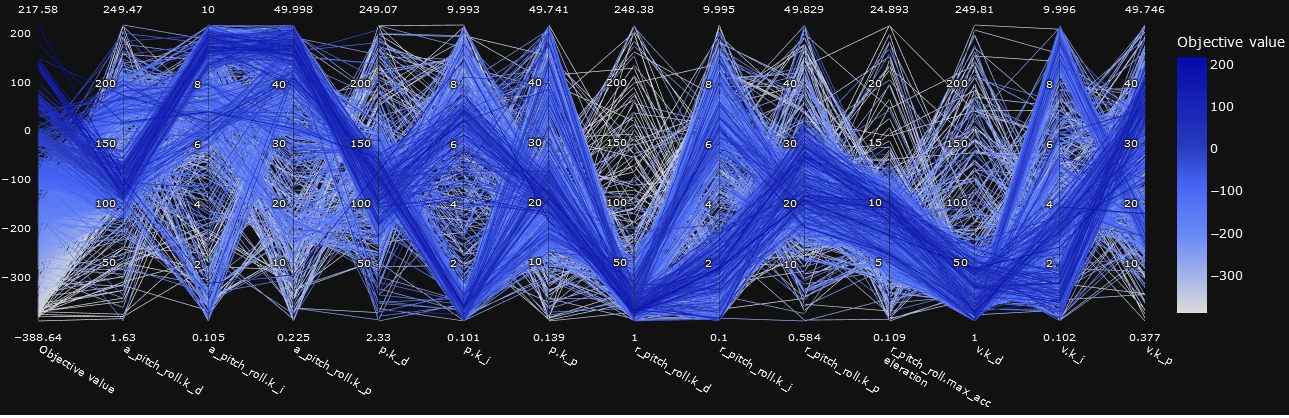}
			\caption{Cascaded PID only.}
			\label{fig:uav_wind_pid_5_90_parallel}
		\end{subfigure}
		\begin{subfigure}{\textwidth}
			\centering
			\includegraphics[width=1.0\textwidth]{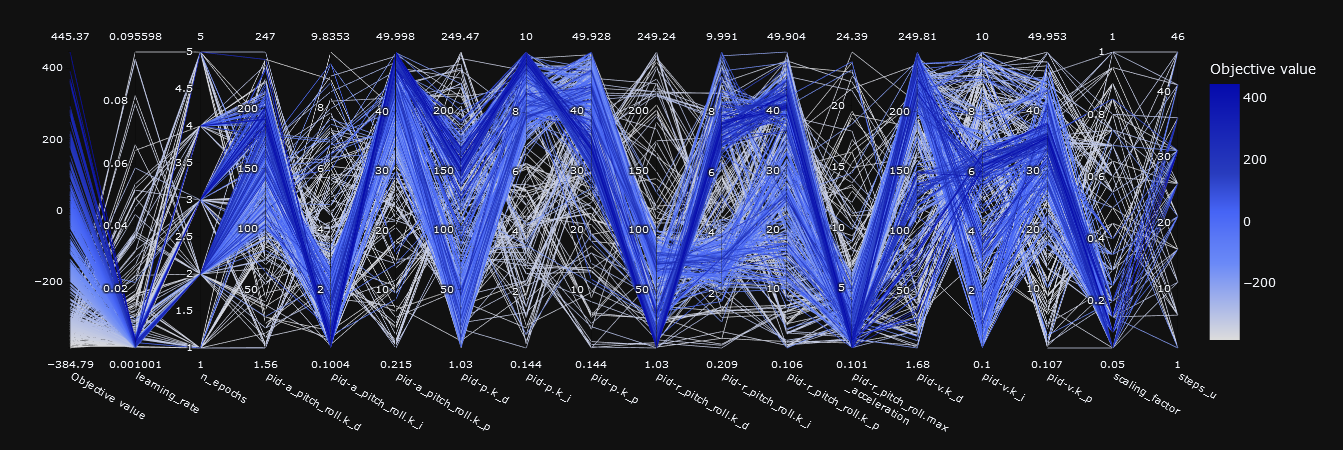}
			\caption{Supervisory RL and cascaded PID.}
			\label{fig:uav_wind_rlpid_5_90_parallel}
		\end{subfigure}
		\caption{A parallel coordinate plot of controller parameter combinations, shaded by mean total reward over the same 10 episodes in wind conditions. The darker track of lines indicates the combinations that performed well on the navigation task.}
	\end{figure}

	A hyperparameter search is run for a PID-only controller for wind disturbances. Figure \ref{fig:uav_wind_pid_5_90_parallel} shows the results of $1000$ hyperparameter optimization trials for one wind condition. Notably, the total reward over 10 test episodes degrades substantially compared to the nominal case with no wind. The average best performance from disturbances from $0\degree$ to $90\degree$ is $317$ for PID, compared to the nominal case of $463$. This is expected since the PID represents reactive control. It can accommodate disturbances, but only to a certain extent without re-tuning all the PID parameters. For the position controller, the most important parameter is $k_d$. For velocity control, it is $k_i$. For attitude control, it is $k_p$. And for rate control, the key parameter is $k_d$. 
	
	\begin{table}[]
		\footnotesize
		\centering
		\begin{tabular}{llrrrrrl}
			\toprule
			&     &  Nominal &  Nominal $\sigma$ /\% &    Wind &  Wind $\sigma$ /\% &  Change /\% &  Notable \\
			Controller & Params &          &                      &         &                   &            &          \\
			\midrule
			Attitude & $k_d$ &    2.182 &              159.562 & 129.924 &            21.353 &   5853.095 &     True \\
			& $k_i$ &    0.490 &               34.428 &   8.026 &            31.822 &   1537.930 &     True \\
			& $k_p$ &   33.099 &                3.528 &  40.167 &            30.672 &     21.353 &    False \\
			Position & $k_d$ &  123.846 &                6.111 & 119.274 &            10.739 &     -3.692 &    False \\
			& $k_i$ &    3.334 &                7.591 &   3.569 &            93.197 &      7.039 &    False \\
			& $k_p$ &   46.091 &                3.602 &  21.704 &            24.494 &    -52.911 &     True \\
			Rate & $k_d$ &   56.596 &               12.557 &  15.397 &            62.921 &    -72.795 &     True \\
			& $k_i$ &    5.238 &                8.657 &   2.649 &            80.416 &    -49.431 &    False \\
			& $k_p$ &   27.040 &                4.462 &  23.469 &            19.329 &    -13.206 &    False \\
			& $max_{acc}$ &    5.376 &               50.992 &   8.478 &            20.002 &     57.717 &     True \\
			Velocity & $k_d$ &   92.076 &                6.026 &  29.712 &            68.644 &    -67.731 &    False \\
			& $k_i$ &    5.039 &                4.172 &   2.725 &            43.418 &    -45.927 &     True \\
			& $k_p$ &   24.393 &                5.736 &  32.050 &            22.363 &     31.393 &     True \\
			\bottomrule
		\end{tabular}
		\caption{Comparison of the top 10 performing parameters for the cascaded PID controller. The ``Notable" column is true for percent changes larger than the standard deviations of the parameters. Parameters for yaw control are set to 0. Altitude velocity and altitude rate controllers are kept as P-only controllers with $k_p=1,10$ respectively.}
		\label{tbl:pid_params}
	\end{table}
	
	% ---------------------------------
	\subsection{Trained robust RL supervisory control under wind}
	% ---------------------------------
	
	In this section, we evaluate the efficacy of the proposed supervisory RL controller with wind disturbances. First, a hyperparameter search consisting of 500 trials is conducted with RL control. Searches are conducted for two prevailing wind disturbances. A parallel plot of hyperparameter combinations is shown in Figure \ref{fig:uav_wind_rlpid_5_90_parallel}. Table \ref{tbl:rlpid_params} shows a comparison of the best parameters for RL control for $5N$ wind coming from $90\degree$ to the direction of flight.
	For the RL controller, Table \ref{tbl:rlpid_params} shows that position and velocity $k_p$ decrease with wind, but $k_i$ increases. The scaling factor for RL actions decreases by 75\%, and the frequency of actions $f_{RL}$ decreases as well. That is, the RL controller is making smaller, less frequent, but ultimately more persistent changes to the waypoint and velocity references. The downstream attitude and rate controllers' higher $k_p$ are quick to act on those changes cascading through the system.
	
	\begin{table}[]
		\footnotesize
		\centering
		\begin{tabular}{llrrrrrl}
			\toprule
			&     &   Nominal &  Nominal $\sigma$ /\% &    Wind &  Wind $\sigma$ /\% &  Change /\% &  Notable \\
			Controller & Params &           &                      &         &                   &            &          \\
			\midrule
			RL  & batch size &   217.600 &                9.301 & 128.000 &             0.000 &    -41.176 &     True \\
			& learning rate &     0.001 &               40.624 &   0.003 &            10.301 &    105.725 &     True \\
			& epochs &     1.200 &               35.136 &   3.300 &            20.453 &    175.000 &     True \\
			& steps & 15372.800 &               40.206 & 825.600 &             6.005 &    -94.629 &     True \\
			& scaling factor &     0.485 &               20.067 &   0.120 &            21.517 &    -75.258 &     True \\
			& steps u &     2.500 &              189.737 &  31.000 &             0.000 &   1140.000 &     True \\
			Attitude & $k_d$ &    54.174 &               32.153 & 200.493 &             9.926 &    270.088 &     True \\
			& $k_i$ &     1.023 &               42.851 &   0.335 &            33.189 &    -67.237 &     True \\
			& $k_p$ &    32.092 &               13.841 &  48.807 &             2.933 &     52.084 &     True \\
			Position & $k_d$ &   203.020 &                8.490 & 156.692 &             5.118 &    -22.819 &     True \\
			& $k_i$ &     6.786 &                6.730 &   9.742 &             1.722 &     43.549 &     True \\
			& $k_p$ &    44.851 &                7.058 &  34.399 &             3.732 &    -23.303 &     True \\
			Rate & $k_d$ &     6.568 &              113.713 &   2.594 &            66.204 &    -60.503 &    False \\
			& $k_i$ &     3.209 &               65.263 &   7.637 &             3.711 &    137.999 &     True \\
			& $k_p$ &    27.769 &               11.064 &  41.355 &             3.028 &     48.928 &     True \\
			& $max_{acc}$ &     0.973 &               52.683 &   0.728 &            72.387 &    -25.120 &    False \\
			Velocity & $k_d$ &   227.655 &                3.126 & 117.124 &            65.095 &    -48.552 &    False \\
			& $k_i$ &     3.404 &               72.105 &   6.187 &             5.269 &     81.772 &     True \\
			& $k_p$ &    47.343 &                3.569 &  36.024 &             3.124 &    -23.909 &     True \\
			\bottomrule
		\end{tabular}
		\caption{Comparison of the top 10 performing parameters for the RL-supervised cascaded PID controller. The ``Notable" column is true for percent changes larger than the standard deviations of the parameters. Parameters for yaw control are set to 0. Altitude velocity and rate controllers are kept as P-only controllers with $k_p=1,10$, respectively.}
		\label{tbl:rlpid_params}
	\end{table}
	
\begin{figure}[ht]
		\centering
		\begin{subfigure}[t]{0.45\textwidth}
			\centering
			\includegraphics[width=\textwidth]{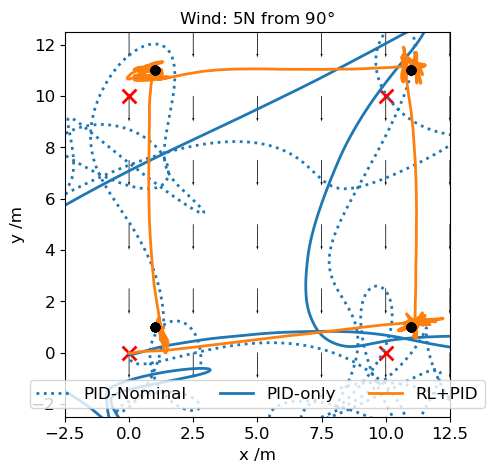}
			\caption{Wind coming from 90$\degree$.}
			\label{fig:uav_wind_rl_vs_pid_5_90}
		\end{subfigure}
		\begin{subfigure}[t]{0.45\textwidth}
			\centering
			\includegraphics[width=\textwidth]{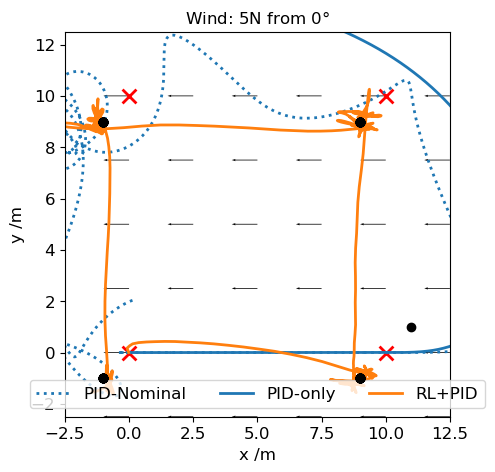}
			\caption{Wind coming from 0$\degree$.}
			\label{fig:uav_wind_rl_vs_pid_5_0}
		\end{subfigure}
		\caption{Performance of navigation under four granular tracking problems stacked together, representing a square trajectory. Each granular problem is navigation to the waypoint marked by ``x".}
		\label{fig:uav_wind_rl_vs_pid}
	\end{figure}

	Figure \ref{fig:uav_wind_rl_vs_pid} shows how the supervisory controller interacts with a navigation problem for two wind conditions. The black points represent the changed waypoints given from the RL supervisor. They are minor changes in position references for the PID controller to follow. Also shown are the trajectories of the nominal PID controller, and the PID controller trained for use with RL but used alone. A nominal PID controller trained on disturbances can complete some trajectories, albeit with overshoots and deviations from the shortest path. When the PID controller trained along with the supervisor is tested alone, it is unable to finish the trajectory.The plots show that a PID controller alone is not sufficient to withstand the simulated disturbance.
	
\begin{figure}[ht]
		\centering
		\begin{subfigure}[t]{0.45\textwidth}
			\centering
			\includegraphics[width=\textwidth]{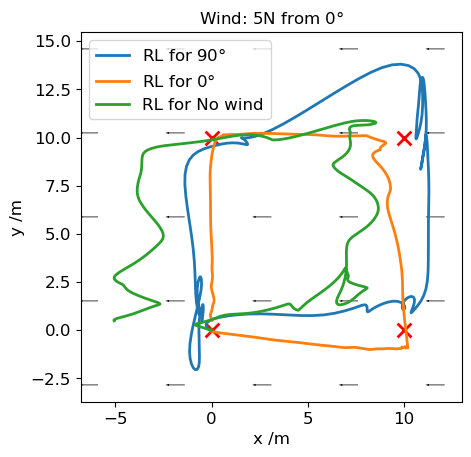}
			\caption{Supervisory RL.}
			\label{fig:uav_wind_rlpid_robustness}
		\end{subfigure}
		\begin{subfigure}[t]{0.65\textwidth}
			\centering
			\includegraphics[width=\textwidth]{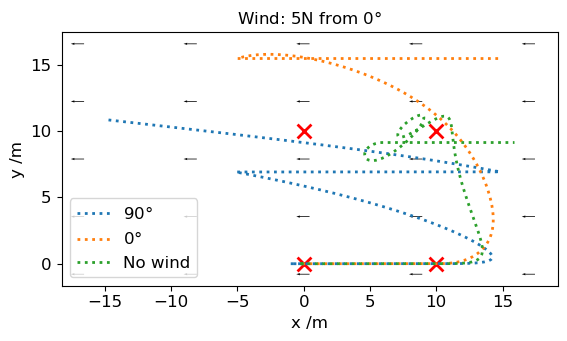}
			\caption{Cascaded PID only.}
			\label{fig:uav_wind_pid_robustness}
		\end{subfigure}
		\caption{A comparison of the robustness of (a) Supervisory RL and (b) PID-only for a wind condition. Controllers optimized for three different disturbances are tested for a single wind condition.}
		\label{fig:uav_wind_rl_vs_pid_robustness}
	\end{figure}

	Figure \ref{fig:uav_wind_rl_vs_pid_robustness} shows RL and PID controllers, optimized for different disturbances, operating on a small trajectory under a single wind condition. While the RL controller is able to reach all waypoints, the PID controllers (both nominal and optimized for wind) fail after reaching one or two waypoints.
	
\begin{figure}[ht]
		\centering
		\begin{subfigure}{0.45\textwidth}
			\centering
			\includegraphics[width=\textwidth]{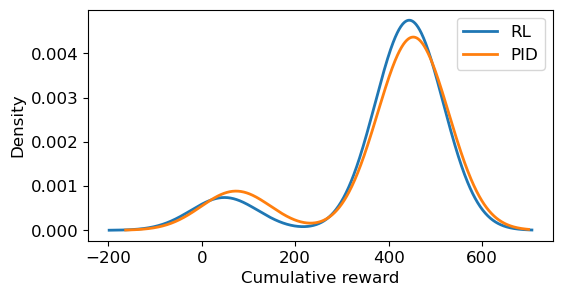}
			\caption{No wind.}
			\label{fig:uav_density_pid_vs_rl}   
		\end{subfigure}
		\begin{subfigure}{0.45\textwidth}
			\includegraphics[width=\textwidth]{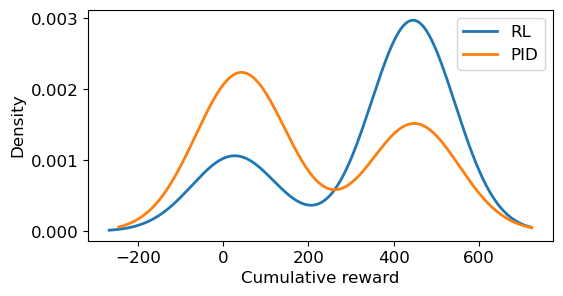}
			\caption{Wind coming from $90\degree$.}
			\label{fig:uav_wind_density_pid_vs_rl_5_90}   
		\end{subfigure}
		\caption{Performance comparison of running 30 experiments for waypoint following using PID-only, and PID with RL waypoint supervision.}
		\label{fig:uav_density_plot}
	\end{figure}

	We generalize this result by evaluating trajectory tracking performance over multiple episodes. For nominal conditions, RL control and PID control produce similar results. However, with wind disturbance, PID-only control degrades, but the RL supervision remains robust. Figure \ref{fig:uav_density_plot} shows a density plot of episodic rewards for nominal and wind conditions for both controllers.
	
\begin{figure}[ht]
		\centering
		\includegraphics[width=0.6\linewidth]{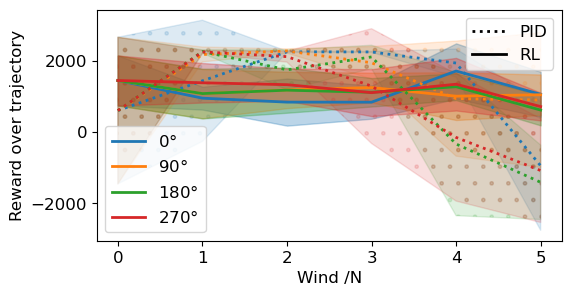}
		\caption{Performance over a square trajectory with increasing wind disturbance. RL and PID controllers trained for $5N$ wind from $0\degree$ are used over 5 trials.}
		\label{fig:uav_rl_vs_pid_degradation}
	\end{figure}

	We conducted another robustness study by varying the wind force magnitude from four directions (see Figure \ref{fig:uav_rl_vs_pid_degradation}). Performance was measured by total rewards over the square trajectories shown previously. Supervised RL controllers could maintain consistent performance with lower variance as the magnitude of the disturbance increased. On the other hand, the PID-only control produced a non-monotonic response with respect to the wind magnitude.
	
	Why does the supervisory RL controller exhibit robustness to disturbances? Both controllers are trained on navigation problems initialized with random velocities and positions. Both have encountered states that may be present in nominal and disturbed environments, such as a net lateral velocity perpendicular to the reference position vector. But, a PID-only controller is unable to continue to plan actions over the trajectory because it needs a reference to follow. From Table \ref{tbl:pid_params}, a PID controller compensates for wind by reducing position $k_p$, and increasing velocity $k_p$. Most significantly, the attitude $k_d, k_i$ terms see an order of magnitude increase. That is, the velocity reference due to a position error is smaller. But, the attitude references respond faster to any change in velocity. And consequently, the angular rate references, determining the torque allocation for lateral motion, are more persistent and damped to counteract the effects of wind drag. While such persistent actions may counteract wind, under nominal conditions they would cause overshoot.
	
	The RL controller, on the other hand, has memorized optimal actions from encountered states during training. When a state is encountered corresponding to a specific wind disturbance, such as a perpendicular wind velocity in comparison to the reference position vector, the controller can recall the appropriate actions. Unlike a PID-only controller, the size of the position error to track given to the PID controller, and the cascaded control logic, varies with the state of the vehicle besides position. For example, for headwinds, the position error is increased to give aggressive velocity references. For tailwinds, the error is decreased to ensure a smaller overshoot.
	
\begin{figure}[ht]
		\centering
		\begin{subfigure}[t]{0.45\textwidth}
			\centering
			\includegraphics[width=\textwidth]{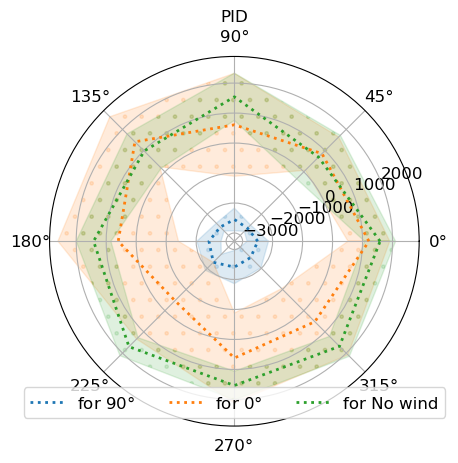}
			\caption{Mean rewards over 10 square trajectories with random initial positions/velocities using PID.}
			\label{fig:uav_polar_pid}
		\end{subfigure}
		\begin{subfigure}[t]{0.45\textwidth}
			\centering
			\includegraphics[width=\textwidth]{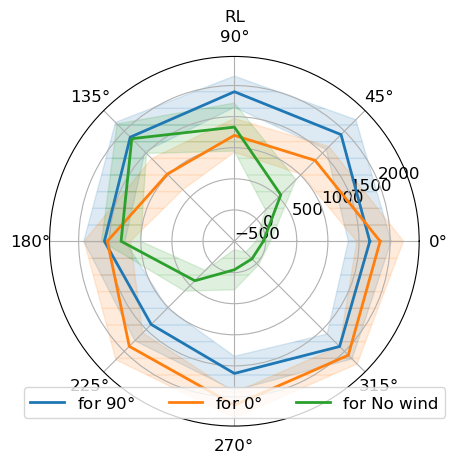}
			\caption{Mean rewards over 10 square trajectories with random initial positions/velocities using RL.}
			\label{fig:uav_polar_rl}
		\end{subfigure}
		\begin{subfigure}[t]{0.44\textwidth}
			\centering
			\includegraphics[width=\textwidth]{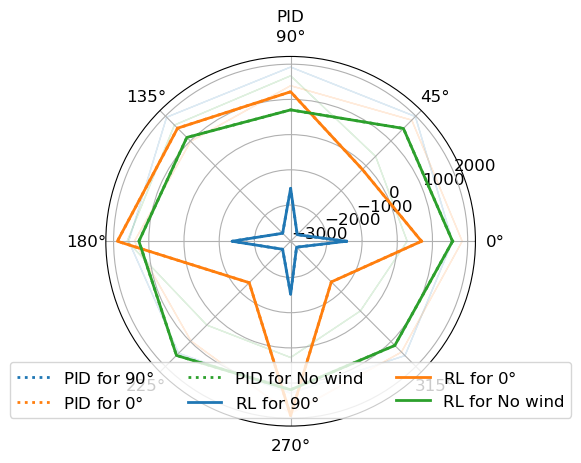}
			\caption{Total rewards over 4 episodes comprising the square trajectory using PID.}
			\label{fig:uav_polar_pid_square}
		\end{subfigure}
		\begin{subfigure}[t]{0.45\textwidth}
			\centering
			\includegraphics[width=\textwidth]{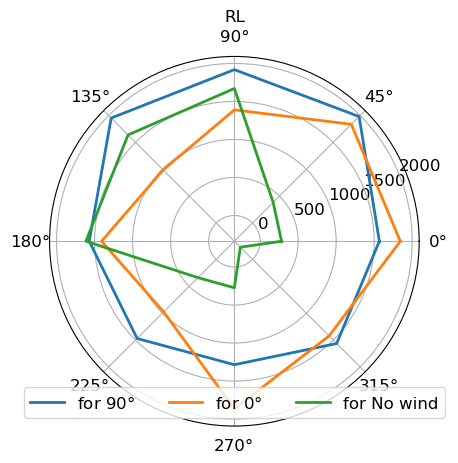}
			\caption{Total rewards over 4 episodes comprising the square trajectory using RL.}
			\label{fig:uav_polar_rl_square}
		\end{subfigure}
		\caption{A polar plot of rewards for different controllers being tested under $5N$ wind coming from various headings.}
		\label{fig:uav_polar_pid_vs_rl_both}
	\end{figure}

	We further reinforce this result using Figure \ref{fig:uav_polar_pid_vs_rl_both}. Controllers, nominal and optimized for wind, are pitted against wind coming from eight different directions and evaluated over 10 episodes each. The PID-only controllers exhibit a larger variance in performance across the disturbances. On the other hand, RL supervision shows more consistent performance under novel disturbances.
	
	Furthermore, the actions of RL controllers under different disturbances are related. The UAV system dynamics exhibit a rotational symmetry on the $x, y$ plane. The state change magnitude in response to forces and torques along the $x$-axis is similar to that along the $y$-axis, except in a different direction. The supervisory controllers exhibit a similar relationship.
	
\begin{figure}[ht]
		\centering
		\begin{subfigure}[t]{0.45\textwidth}
			\centering
			\includegraphics[width=\textwidth]{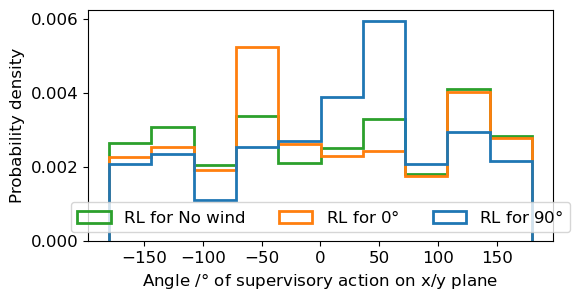}
			\caption{Histogram of RL controller actions' angles}
			\label{fig:uav_rl_angles}
		\end{subfigure}
		\begin{subfigure}[t]{0.45\textwidth}
			\centering
			\includegraphics[width=\textwidth]{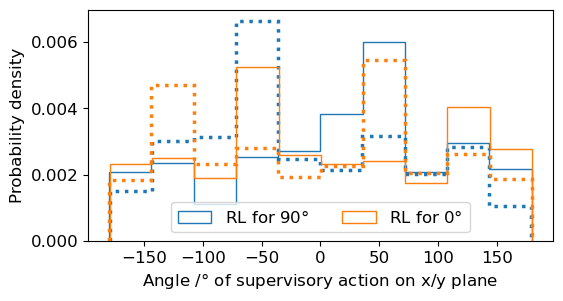}
			\caption{Actions' angles shifted by 90$\degree$}
			\label{fig:uav_rl_angles_xformed}
		\end{subfigure}
		\begin{subfigure}[t]{0.3\textwidth}
			\centering
			\includegraphics[width=\textwidth]{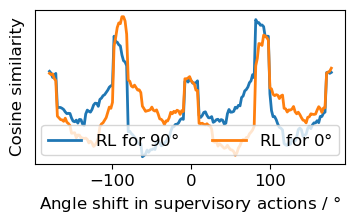}
			\caption{Cosine similarity between histograms of angles of supervisory actions as they are shifted, against the histogram of the other controller's actions' angles.}
			\label{fig:uav_rl_angles_cosine}
		\end{subfigure}
		\caption{A density histogram showing relationship of the angles of the position vectors of the supervisory actions to modify the reference waypoint of the PID controller, calculated over 1000 randomly generated input states.}
		\label{fig:uav_rl_angles_all}
	\end{figure}

	Figure \ref{fig:uav_rl_angles} is a histogram of the angles of supervisory action in the $x,y$-plane, relative to the positive $x$-axis. Figure \ref{fig:uav_rl_angles_xformed} is the same histogram, but shifted by $90\degree$, representing the change in heading of the wind disturbances for respective controllers. We note that the modes of the distributions align. Further, the cosine similarity (Figure \ref{fig:uav_rl_angles_cosine}), between histograms is the highest for a $90\degree$ translation among all other angles. The relationship between the actions that the disturbance conditions point to presents an interesting topic for future research. By identifying such relationships, a nominal controller can be adapted to a new control task without the need for tuning from scratch.
	
	\subsection{Operating robust RL control on longer trajectories}
	
\begin{figure}[ht]
		\centering
		\begin{subfigure}{0.45\textwidth}
			\centering \includegraphics[width=\textwidth]{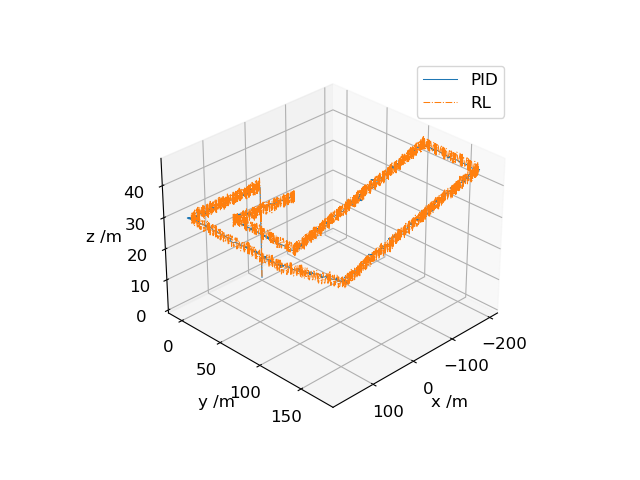}
			\caption{Controllers for no wind, tested on no wind disturbances.}
			\label{fig:uav_long_rl_vs_pid}
		\end{subfigure}
		\begin{subfigure}{0.45\textwidth}
			\centering \includegraphics[width=\textwidth]{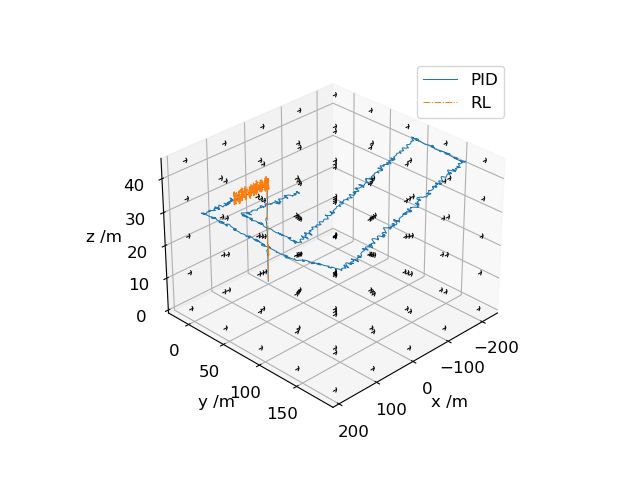}
			\caption{Controllers for no wind, tested on smaller $3N$ wind coming from $0\degree$.}
			\label{fig:uav_windweak_long_rl_vs_pid_fail}
		\end{subfigure}
		\begin{subfigure}{0.45\textwidth}
			\centering \includegraphics[width=\textwidth]{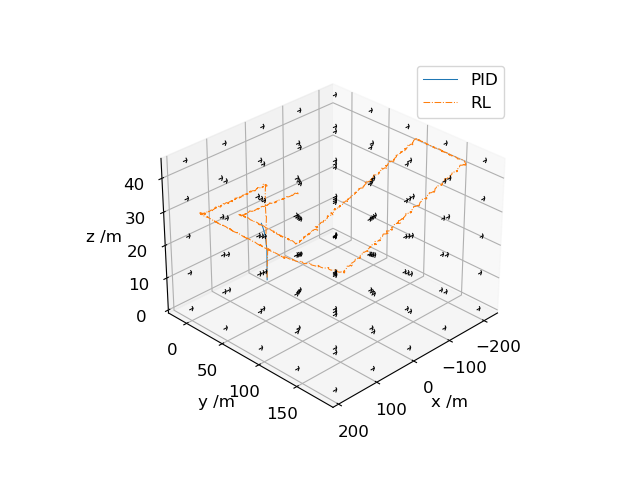}
			\caption{Controllers for $5N$ wind from $90\degree$, tested on $0\degree$ wind.}
			\label{fig:uav_wind_long_rl_vs_pid}
		\end{subfigure}
		\begin{subfigure}{0.45\textwidth}
			\centering \includegraphics[width=\textwidth]{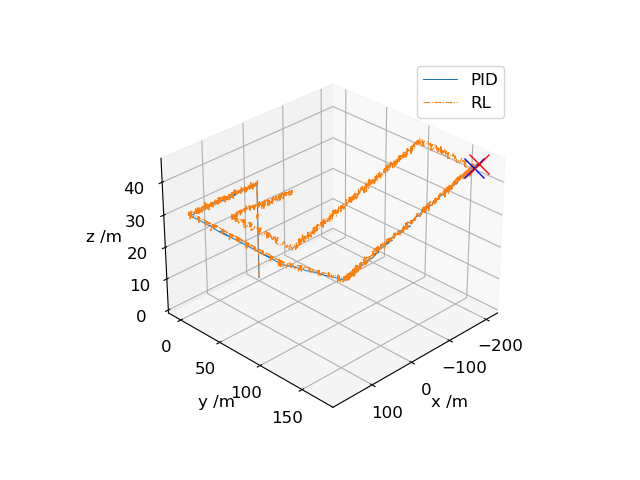}
			\caption{RL controllers for $5N$ wind from $90\degree$ and nominal PID controller, tested on $0\degree$ wind starting in the middle of the flight (blue cross). The cascaded PID-only controller fails shortly after (red cross).}
			\label{fig:uav_long_rl_vs_pid_middle}
		\end{subfigure}
		\caption{Simulations of UAV flight under different disturbances over a longer flight.}
		\label{fig:uav_wind_rl_vs_pid_long}
		
	\end{figure}

	Figure \ref{fig:uav_wind_rl_vs_pid_long} simulates longer trajectories by solving granular navigation problems in sequence as described in Algorithm \ref{alg:uav_rl_operation}. The trajectories are simulations of real-world experiments for the Tarot T-18 octo-rotor conducted by NASA Langley Research Center. The experiments are conducted to show the generalizability of our proposed approach. When there are no disturbances, both PID and RL controllers finish the designated path (see Figure \ref{fig:uav_long_rl_vs_pid}). However, under wind disturbance, when an RL controller, not trained on wind conditions is run, it fails (see Figure \ref{fig:uav_windweak_long_rl_vs_pid_fail}). But when an RL controller trained on one wind disturbance encounters another novel disturbance, it is able to perform well (see Figure \ref{fig:uav_wind_long_rl_vs_pid}). Furthermore, the RL controller can operate in nominal and windy conditions that may occur  in the same flight. On the other hand, a PID controller fails early when it encounters a disturbance (see Figure \ref{fig:uav_long_rl_vs_pid_middle}). These results reinforce our hypothesis that the robust trajectory control problem can be decomposed into smaller sub-problems that are more tractable to solve for RL. At the same time, the controller can maintain generalizability for longer, more realistic trajectories.
	
	% ======================================================
	
	\section{Conclusions and Future work}\label{sec:conclusion}
	
	In this work, we presented a supervisory control architecture for a class of unmanned aerial vehicles that is robust to wind disturbances. The supervisory RL controller is based on optimizing time to waypoint and deviations from the shortest path. The RL supervisor breaks down a larger trajectory into granular navigation problems and modifies the position references of the underlying PID controller.
	
	Experiments with an octo-rotor simulation showed that, for wind disturbances, both RL and PID controllers tend to prefer persistent actions to maximally counteract the drag forces. An RL controller, being proactive, can modify the position reference marginally, causing the underlying cascaded PID logic to transparently cancel out disturbance effects.
	
	Another characteristic of RL is robustness to disturbances. The RL controller  \textit{memorizes} the value of actions in the form of a neural network that operates over a continuous domain of states. Some states may occur during operation under the controller's nominal environment as well as for novel disturbances. Therefore, the controller can recall (or extrapolate) the most valuable action for a state. Contrast this with another planning algorithm, Model-Predictive Control, which will have to simulate a model recurrently to derive the best action. Under a new disturbance, the model may be inaccurate and thus output sub-optimal actions \citep{ahmed2018comparison}.
	
	Finally, we explained the robustness of RL control by illustrating how changes in control actions and disturbances are related. By understanding the transformation in the environment (and system dynamics), an adaptive control strategy can be developed that will quickly respond to changes in the control problem by transforming the control policy correspondingly.
	
	Prior work with RL and UAVs have made simplifying assumptions about control. For example, relaxed time constraints for inferring RL actions, or access to all actuators on the vehicle. While these assumptions can be realized with powerful computational hardware and bespoke circuitry, they do not lend themselves to wider, more general adoption. Our proposed supervisory approach is interoperable with popular flight control software like the  Ardupilot. By keeping the RL control supervisory, we can separate it from the low-level control application, mitigating risks of complete software failure. The system necessarily operates at lower frequencies than the low-level PID control. Therefore, it can account for complex neural networks to learn RL control and it can operate on a wider range of computers. In addition, it can run on a separate (potentially remote) device, lowering payload requirements (battery, computer) for the vehicle.
	
	\backmatter
	
	\section*{Declarations}
	
	\subsection{Funding}
	This research was funded by NASA  grant number 80NSSC21M0087.
	
	\subsection{Competing interests}
	
	The authors declare they have no competing interests.
	
	\subsection{Authors' contributions}
	
	Conceptualization, Ahmed, I., Quinones-Grueiro, M., and Biswas, G.; methodology, Ahmed, I., Quinones-Grueiro, M., and Biswas, G.; software, Ahmed, I.; validation, Quinones-Grueiro, M., and Biswas, G.; formal analysis, Ahmed, I.; investigation, Ahmed, I.; data curation, Ahmed, I.; writing---original draft preparation, Ahmed, I.; writing---review and editing, Ahmed, I., Quinones-Grueiro, M., and Biswas, G.; visualization, Ahmed, I.; supervision, Biswas, G.;
	
	\subsection{Code availability}
	
	The code and data for this work are available at \url{https://git.isis.vanderbilt.edu/ahmedi/transfer_similarity}.
	
	\subsection{Ethics approval}
	Not applicable.
	\subsection{Consent to participate}
	Not applicable.
	\subsection{Consent for publication}
	Not applicable.
	
	\begin{appendices}
		
		\section{Implementation details}\label{sec:appendix}
		
		The code and data for this work are available at \url{https://git.isis.vanderbilt.edu/ahmedi/transfer_similarity}.
		
		The testbed is a Tarot T-18 octorotor simulation, using the multirotor python library \url{https://github.com/hazrmard/multirotor} \citep{ahmed2022high}. The simulation time step is 0.01s. Attitude and rate controllers run at 100Hz, but the position and velocity controllers run at 10Hz. Transient dynamics of the motors are ignored when applying propeller speed signals to the UAV frame.
		
		All reinforcement learning experiments were carried out using the \texttt{StableBaselines3} python library (\url{https://stable-baselines3.readthedocs.io/en/master/}). Proximal Policy Optimization was used for RL, with default parameters, except when overridden by the hyperparameter search documented in this work. The RL controller used a policy and value network of width 128, and depth 2 with $\tanh$ activations.
		
		Hyperparameter optimization uses the Optuna library for finding PID and RL parameters (\url{https://optuna.readthedocs.io/en/stable/index.html})\citep{optuna_2019}.
		
	\end{appendices}
	
	%%===========================================================================================%%
	%% If you are submitting to one of the Nature Portfolio journals, using the eJP submission   %%
	%% system, please include the references within the manuscript file itself. You may do this  %%
	%% by copying the reference list from your .bbl file, paste it into the main manuscript .tex %%
	%% file, and delete the associated \verb+\bibliography+ commands.                            %%
	%%===========================================================================================%%
	
	\bibliography{ref/litreview,ref/ref}% common bib file

%% BioMed_Central_Bib_Style_v1.01

\begin{thebibliography}{40}
% BibTex style file: bmc-mathphys.bst (version 2.1), 2014-07-24
\ifx \bisbn   \undefined \def \bisbn  #1{ISBN #1}\fi
\ifx \binits  \undefined \def \binits#1{#1}\fi
\ifx \bauthor  \undefined \def \bauthor#1{#1}\fi
\ifx \batitle  \undefined \def \batitle#1{#1}\fi
\ifx \bjtitle  \undefined \def \bjtitle#1{#1}\fi
\ifx \bvolume  \undefined \def \bvolume#1{\textbf{#1}}\fi
\ifx \byear  \undefined \def \byear#1{#1}\fi
\ifx \bissue  \undefined \def \bissue#1{#1}\fi
\ifx \bfpage  \undefined \def \bfpage#1{#1}\fi
\ifx \blpage  \undefined \def \blpage #1{#1}\fi
\ifx \burl  \undefined \def \burl#1{\textsf{#1}}\fi
\ifx \doiurl  \undefined \def \doiurl#1{\url{https://doi.org/#1}}\fi
\ifx \betal  \undefined \def \betal{\textit{et al.}}\fi
\ifx \binstitute  \undefined \def \binstitute#1{#1}\fi
\ifx \binstitutionaled  \undefined \def \binstitutionaled#1{#1}\fi
\ifx \bctitle  \undefined \def \bctitle#1{#1}\fi
\ifx \beditor  \undefined \def \beditor#1{#1}\fi
\ifx \bpublisher  \undefined \def \bpublisher#1{#1}\fi
\ifx \bbtitle  \undefined \def \bbtitle#1{#1}\fi
\ifx \bedition  \undefined \def \bedition#1{#1}\fi
\ifx \bseriesno  \undefined \def \bseriesno#1{#1}\fi
\ifx \blocation  \undefined \def \blocation#1{#1}\fi
\ifx \bsertitle  \undefined \def \bsertitle#1{#1}\fi
\ifx \bsnm \undefined \def \bsnm#1{#1}\fi
\ifx \bsuffix \undefined \def \bsuffix#1{#1}\fi
\ifx \bparticle \undefined \def \bparticle#1{#1}\fi
\ifx \barticle \undefined \def \barticle#1{#1}\fi
\bibcommenthead
\ifx \bconfdate \undefined \def \bconfdate #1{#1}\fi
\ifx \botherref \undefined \def \botherref #1{#1}\fi
\ifx \url \undefined \def \url#1{\textsf{#1}}\fi
\ifx \bchapter \undefined \def \bchapter#1{#1}\fi
\ifx \bbook \undefined \def \bbook#1{#1}\fi
\ifx \bcomment \undefined \def \bcomment#1{#1}\fi
\ifx \oauthor \undefined \def \oauthor#1{#1}\fi
\ifx \citeauthoryear \undefined \def \citeauthoryear#1{#1}\fi
\ifx \endbibitem  \undefined \def \endbibitem {}\fi
\ifx \bconflocation  \undefined \def \bconflocation#1{#1}\fi
\ifx \arxivurl  \undefined \def \arxivurl#1{\textsf{#1}}\fi
\csname PreBibitemsHook\endcsname

%%% 1
\bibitem[\protect\citeauthoryear{Cohn et~al.}{2017}]{cohn2017commercial}
\begin{botherref}
\oauthor{\bsnm{Cohn}, \binits{P.}},
\oauthor{\bsnm{Green}, \binits{A.}},
\oauthor{\bsnm{Langstaff}, \binits{M.}},
\oauthor{\bsnm{Roller}, \binits{M.}}:
Commercial drones are here: The future of unmanned aerial systems.
McKinsey \& Company
(2017)
\end{botherref}
\endbibitem

%%% 2
\bibitem[\protect\citeauthoryear{IV et~al.}{2006}]{iv2006assessment}
\begin{barticle}
\bauthor{\bsnm{IV}, \binits{G.P.J.}},
\bauthor{\bsnm{Pearlstine}, \binits{L.G.}},
\bauthor{\bsnm{Percival}, \binits{H.F.}}:
\batitle{An assessment of small unmanned aerial vehicles for wildlife
  research}.
\bjtitle{Wildlife society bulletin}
\bvolume{34}(\bissue{3}),
\bfpage{750}--\blpage{758}
(\byear{2006})
\end{barticle}
\endbibitem

%%% 3
\bibitem[\protect\citeauthoryear{Nelson et~al.}{2019}]{nelson2019view}
\begin{barticle}
\bauthor{\bsnm{Nelson}, \binits{J.R.}},
\bauthor{\bsnm{Grubesic}, \binits{T.H.}},
\bauthor{\bsnm{Wallace}, \binits{D.}},
\bauthor{\bsnm{Chamberlain}, \binits{A.W.}}:
\batitle{The view from above: A survey of the public’s perception of unmanned
  aerial vehicles and privacy}.
\bjtitle{Journal of urban technology}
\bvolume{26}(\bissue{1}),
\bfpage{83}--\blpage{105}
(\byear{2019})
\end{barticle}
\endbibitem

%%% 4
\bibitem[\protect\citeauthoryear{Dhulipalla
  et~al.}{2022}]{dhulipalla2022comparative}
\begin{bchapter}
\bauthor{\bsnm{Dhulipalla}, \binits{A.}},
\bauthor{\bsnm{Han}, \binits{N.}},
\bauthor{\bsnm{Hu}, \binits{H.}},
\bauthor{\bsnm{Hu}, \binits{H.}}:
\bctitle{A comparative study to characterize the effects of adverse weathers on
  the flight performance of an unmanned-aerial-system}.
In: \bbtitle{AIAA AVIATION 2022 Forum},
p. \bfpage{3962}
(\byear{2022})
\end{bchapter}
\endbibitem

%%% 5
\bibitem[\protect\citeauthoryear{Grishin et~al.}{2020}]{grishin2020methods}
\begin{bchapter}
\bauthor{\bsnm{Grishin}, \binits{I.}},
\bauthor{\bsnm{Vishnevsky}, \binits{V.}},
\bauthor{\bsnm{Dinh}, \binits{T.D.}},
\bauthor{\bsnm{Vybornova}, \binits{A.}},
\bauthor{\bsnm{Kirichek}, \binits{R.}}:
\bctitle{Methods for correcting positions of tethered uavs in adverse weather
  conditions}.
In: \bbtitle{2020 12th International Congress on Ultra Modern
  Telecommunications and Control Systems and Workshops (ICUMT)},
pp. \bfpage{308}--\blpage{312}
(\byear{2020}).
\bcomment{IEEE}
\end{bchapter}
\endbibitem

%%% 6
\bibitem[\protect\citeauthoryear{Ahmed et~al.}{2023}]{ahmed2023adaptive}
\begin{bchapter}
\bauthor{\bsnm{Ahmed}, \binits{I.}},
\bauthor{\bsnm{Quinones-Grueiro}, \binits{M.}},
\bauthor{\bsnm{Biswas}, \binits{G.}}:
\bctitle{Adaptive fault-tolerant control of octo-rotor uav under motor faults
  in adverse wind conditions}.
In: \bbtitle{AIAA SCITECH 2023 Forum},
p. \bfpage{2535}
(\byear{2023})
\end{bchapter}
\endbibitem

%%% 7
\bibitem[\protect\citeauthoryear{Ebeid et~al.}{2017}]{ebeid2017survey}
\begin{bchapter}
\bauthor{\bsnm{Ebeid}, \binits{E.}},
\bauthor{\bsnm{Skriver}, \binits{M.}},
\bauthor{\bsnm{Jin}, \binits{J.}}:
\bctitle{A survey on open-source flight control platforms of unmanned aerial
  vehicle}.
In: \bbtitle{2017 Euromicro Conference on Digital System Design (dsd)},
pp. \bfpage{396}--\blpage{402}
(\byear{2017}).
\bcomment{IEEE}
\end{bchapter}
\endbibitem

%%% 8
\bibitem[\protect\citeauthoryear{Bianco et~al.}{2018}]{bianco2018benchmark}
\begin{barticle}
\bauthor{\bsnm{Bianco}, \binits{S.}},
\bauthor{\bsnm{Cadene}, \binits{R.}},
\bauthor{\bsnm{Celona}, \binits{L.}},
\bauthor{\bsnm{Napoletano}, \binits{P.}}:
\batitle{Benchmark analysis of representative deep neural network
  architectures}.
\bjtitle{IEEE access}
\bvolume{6},
\bfpage{64270}--\blpage{64277}
(\byear{2018})
\end{barticle}
\endbibitem

%%% 9
\bibitem[\protect\citeauthoryear{{\AA}str{\"o}m and
  Wittenmark}{2013}]{aastrom2013adaptive}
\begin{bbook}
\bauthor{\bsnm{{\AA}str{\"o}m}, \binits{K.J.}},
\bauthor{\bsnm{Wittenmark}, \binits{B.}}:
\bbtitle{Adaptive Control}.
\bpublisher{Courier Corporation}, \blocation{???}
(\byear{2013})
\end{bbook}
\endbibitem

%%% 10
\bibitem[\protect\citeauthoryear{Adetola et~al.}{2009}]{adetola2009adaptive}
\begin{barticle}
\bauthor{\bsnm{Adetola}, \binits{V.}},
\bauthor{\bsnm{DeHaan}, \binits{D.}},
\bauthor{\bsnm{Guay}, \binits{M.}}:
\batitle{Adaptive model predictive control for constrained nonlinear systems}.
\bjtitle{Systems \& Control Letters}
\bvolume{58}(\bissue{5}),
\bfpage{320}--\blpage{326}
(\byear{2009})
\end{barticle}
\endbibitem

%%% 11
\bibitem[\protect\citeauthoryear{Bemporad et~al.}{2002}]{bemporad2002explicit}
\begin{barticle}
\bauthor{\bsnm{Bemporad}, \binits{A.}},
\bauthor{\bsnm{Morari}, \binits{M.}},
\bauthor{\bsnm{Dua}, \binits{V.}},
\bauthor{\bsnm{Pistikopoulos}, \binits{E.N.}}:
\batitle{The explicit linear quadratic regulator for constrained systems}.
\bjtitle{Automatica}
\bvolume{38}(\bissue{1}),
\bfpage{3}--\blpage{20}
(\byear{2002})
\end{barticle}
\endbibitem

%%% 12
\bibitem[\protect\citeauthoryear{Anderson et~al.}{1988}]{anderson1988rule}
\begin{bchapter}
\bauthor{\bsnm{Anderson}, \binits{K.L.}},
\bauthor{\bsnm{Blankenship}, \binits{G.L.}},
\bauthor{\bsnm{Lebow}, \binits{L.G.}}:
\bctitle{A rule-based adaptive pid controller}.
In: \bbtitle{Proceedings of the 27th IEEE Conference on Decision and Control},
pp. \bfpage{564}--\blpage{569}
(\byear{1988}).
\bcomment{IEEE}
\end{bchapter}
\endbibitem

%%% 13
\bibitem[\protect\citeauthoryear{Tjokro and Shah}{1985}]{tjokro1985adaptive}
\begin{bchapter}
\bauthor{\bsnm{Tjokro}, \binits{S.}},
\bauthor{\bsnm{Shah}, \binits{S.L.}}:
\bctitle{Adaptive pid control}.
In: \bbtitle{1985 American Control Conference},
pp. \bfpage{1528}--\blpage{1534}
(\byear{1985}).
\bcomment{IEEE}
\end{bchapter}
\endbibitem

%%% 14
\bibitem[\protect\citeauthoryear{Sadeghzadeh
  et~al.}{2012}]{sadeghzadeh2012active}
\begin{bchapter}
\bauthor{\bsnm{Sadeghzadeh}, \binits{I.}},
\bauthor{\bsnm{Mehta}, \binits{A.}},
\bauthor{\bsnm{Chamseddine}, \binits{A.}},
\bauthor{\bsnm{Zhang}, \binits{Y.}}:
\bctitle{Active fault tolerant control of a quadrotor uav based on
  gainscheduled pid control}.
In: \bbtitle{2012 25th IEEE Canadian Conference on Electrical and Computer
  Engineering (CCECE)},
pp. \bfpage{1}--\blpage{4}
(\byear{2012}).
\bcomment{IEEE}
\end{bchapter}
\endbibitem

%%% 15
\bibitem[\protect\citeauthoryear{Pounds et~al.}{2012}]{pounds2012stability}
\begin{barticle}
\bauthor{\bsnm{Pounds}, \binits{P.E.}},
\bauthor{\bsnm{Bersak}, \binits{D.R.}},
\bauthor{\bsnm{Dollar}, \binits{A.M.}}:
\batitle{Stability of small-scale uav helicopters and quadrotors with added
  payload mass under pid control}.
\bjtitle{Autonomous Robots}
\bvolume{33}(\bissue{1}),
\bfpage{129}--\blpage{142}
(\byear{2012})
\end{barticle}
\endbibitem

%%% 16
\bibitem[\protect\citeauthoryear{Marks et~al.}{2012}]{marks2012control}
\begin{bchapter}
\bauthor{\bsnm{Marks}, \binits{A.}},
\bauthor{\bsnm{Whidborne}, \binits{J.F.}},
\bauthor{\bsnm{Yamamoto}, \binits{I.}}:
\bctitle{Control allocation for fault tolerant control of a vtol octorotor}.
In: \bbtitle{Proceedings of 2012 UKACC International Conference on Control},
pp. \bfpage{357}--\blpage{362}
(\byear{2012}).
\bcomment{IEEE}
\end{bchapter}
\endbibitem

%%% 17
\bibitem[\protect\citeauthoryear{Zulu et~al.}{2014}]{zulu2014review}
\begin{barticle}
\bauthor{\bsnm{Zulu}, \binits{A.}},
\bauthor{\bsnm{John}, \binits{S.}}, \betal:
\batitle{A review of control algorithms for autonomous quadrotors}.
\bjtitle{Open Journal of Applied Sciences}
\bvolume{4}(\bissue{14}),
\bfpage{547}
(\byear{2014})
\end{barticle}
\endbibitem

%%% 18
\bibitem[\protect\citeauthoryear{Yu et~al.}{2015}]{yu2015mpc}
\begin{bchapter}
\bauthor{\bsnm{Yu}, \binits{B.}},
\bauthor{\bsnm{Zhang}, \binits{Y.}},
\bauthor{\bsnm{Qu}, \binits{Y.}}:
\bctitle{Mpc-based ftc with fdd against actuator faults of uavs}.
In: \bbtitle{2015 15th International Conference on Control, Automation and
  Systems (ICCAS)},
pp. \bfpage{225}--\blpage{230}
(\byear{2015}).
\bcomment{IEEE}
\end{bchapter}
\endbibitem

%%% 19
\bibitem[\protect\citeauthoryear{Saied et~al.}{2015}]{saied2015fault}
\begin{bchapter}
\bauthor{\bsnm{Saied}, \binits{M.}},
\bauthor{\bsnm{Lussier}, \binits{B.}},
\bauthor{\bsnm{Fantoni}, \binits{I.}},
\bauthor{\bsnm{Francis}, \binits{C.}},
\bauthor{\bsnm{Shraim}, \binits{H.}},
\bauthor{\bsnm{Sanahuja}, \binits{G.}}:
\bctitle{Fault diagnosis and fault-tolerant control strategy for rotor failure
  in an octorotor}.
In: \bbtitle{2015 IEEE International Conference on Robotics and Automation
  (ICRA)},
pp. \bfpage{5266}--\blpage{5271}
(\byear{2015}).
\bcomment{IEEE}
\end{bchapter}
\endbibitem

%%% 20
\bibitem[\protect\citeauthoryear{Pereida and
  Schoellig}{2018}]{pereida2018adaptive}
\begin{bchapter}
\bauthor{\bsnm{Pereida}, \binits{K.}},
\bauthor{\bsnm{Schoellig}, \binits{A.P.}}:
\bctitle{Adaptive model predictive control for high-accuracy trajectory
  tracking in changing conditions}.
In: \bbtitle{2018 IEEE/RSJ International Conference on Intelligent Robots and
  Systems (IROS)},
pp. \bfpage{7831}--\blpage{7837}
(\byear{2018}).
\bcomment{IEEE}
\end{bchapter}
\endbibitem

%%% 21
\bibitem[\protect\citeauthoryear{Chowdhary
  et~al.}{2013}]{chowdhary2013concurrent}
\begin{bchapter}
\bauthor{\bsnm{Chowdhary}, \binits{G.}},
\bauthor{\bsnm{M{\"u}hlegg}, \binits{M.}},
\bauthor{\bsnm{How}, \binits{J.P.}},
\bauthor{\bsnm{Holzapfel}, \binits{F.}}:
\bctitle{Concurrent learning adaptive model predictive control}.
In: \bbtitle{Advances in Aerospace Guidance, Navigation and Control},
pp. \bfpage{29}--\blpage{47}.
\bpublisher{Springer}, \blocation{???}
(\byear{2013})
\end{bchapter}
\endbibitem

%%% 22
\bibitem[\protect\citeauthoryear{Razmi and Afshinfar}{2019}]{razmi2019neural}
\begin{barticle}
\bauthor{\bsnm{Razmi}, \binits{H.}},
\bauthor{\bsnm{Afshinfar}, \binits{S.}}:
\batitle{Neural network-based adaptive sliding mode control design for position
  and attitude control of a quadrotor uav}.
\bjtitle{Aerospace Science and technology}
\bvolume{91},
\bfpage{12}--\blpage{27}
(\byear{2019})
\end{barticle}
\endbibitem

%%% 23
\bibitem[\protect\citeauthoryear{Hamadi et~al.}{2020}]{hamadi2020comparative}
\begin{barticle}
\bauthor{\bsnm{Hamadi}, \binits{H.}},
\bauthor{\bsnm{Lussier}, \binits{B.}},
\bauthor{\bsnm{Fantoni}, \binits{I.}},
\bauthor{\bsnm{Francis}, \binits{C.}},
\bauthor{\bsnm{Shraim}, \binits{H.}}:
\batitle{Comparative study of self tuning, adaptive and multiplexing ftc
  strategies for successive failures in an octorotor uav}.
\bjtitle{Robotics and Autonomous Systems}
\bvolume{133},
\bfpage{103602}
(\byear{2020})
\end{barticle}
\endbibitem

%%% 24
\bibitem[\protect\citeauthoryear{Baldi et~al.}{2022}]{baldi2022ardupilot}
\begin{barticle}
\bauthor{\bsnm{Baldi}, \binits{S.}},
\bauthor{\bsnm{Sun}, \binits{D.}},
\bauthor{\bsnm{Xia}, \binits{X.}},
\bauthor{\bsnm{Zhou}, \binits{G.}},
\bauthor{\bsnm{Liu}, \binits{D.}}:
\batitle{Ardupilot-based adaptive autopilot: architecture and
  software-in-the-loop experiments}.
\bjtitle{IEEE Transactions on Aerospace and Electronic Systems}
\bvolume{58}(\bissue{5}),
\bfpage{4473}--\blpage{4485}
(\byear{2022})
\end{barticle}
\endbibitem

%%% 25
\bibitem[\protect\citeauthoryear{Li et~al.}{2022}]{li2022embedding}
\begin{bchapter}
\bauthor{\bsnm{Li}, \binits{P.}},
\bauthor{\bsnm{Liu}, \binits{D.}},
\bauthor{\bsnm{Xia}, \binits{X.}},
\bauthor{\bsnm{Baldi}, \binits{S.}}:
\bctitle{Embedding adaptive features in the ardupilot control architecture for
  unmanned aerial vehicles}.
In: \bbtitle{2022 IEEE 61st Conference on Decision and Control (CDC)},
pp. \bfpage{3773}--\blpage{3780}
(\byear{2022}).
\bcomment{IEEE}
\end{bchapter}
\endbibitem

%%% 26
\bibitem[\protect\citeauthoryear{Chamseddine
  et~al.}{2015}]{chamseddine2015active}
\begin{barticle}
\bauthor{\bsnm{Chamseddine}, \binits{A.}},
\bauthor{\bsnm{Theilliol}, \binits{D.}},
\bauthor{\bsnm{Zhang}, \binits{Y.}},
\bauthor{\bsnm{Join}, \binits{C.}},
\bauthor{\bsnm{Rabbath}, \binits{C.-A.}}:
\batitle{Active fault-tolerant control system design with trajectory
  re-planning against actuator faults and saturation: application to a
  quadrotor unmanned aerial vehicle}.
\bjtitle{International Journal of Adaptive Control and Signal Processing}
\bvolume{29}(\bissue{1}),
\bfpage{1}--\blpage{23}
(\byear{2015})
\end{barticle}
\endbibitem

%%% 27
\bibitem[\protect\citeauthoryear{Park et~al.}{2013}]{park2013fault}
\begin{barticle}
\bauthor{\bsnm{Park}, \binits{S.}},
\bauthor{\bsnm{Bae}, \binits{J.}},
\bauthor{\bsnm{Kim}, \binits{Y.}},
\bauthor{\bsnm{Kim}, \binits{S.}}:
\batitle{Fault tolerant flight control system for the tilt-rotor uav}.
\bjtitle{Journal of the Franklin Institute}
\bvolume{350}(\bissue{9}),
\bfpage{2535}--\blpage{2559}
(\byear{2013})
\end{barticle}
\endbibitem

%%% 28
\bibitem[\protect\citeauthoryear{Schweighofer and
  Doya}{2003}]{schweighofer2003meta}
\begin{barticle}
\bauthor{\bsnm{Schweighofer}, \binits{N.}},
\bauthor{\bsnm{Doya}, \binits{K.}}:
\batitle{Meta-learning in reinforcement learning}.
\bjtitle{Neural Networks}
\bvolume{16}(\bissue{1}),
\bfpage{5}--\blpage{9}
(\byear{2003})
\end{barticle}
\endbibitem

%%% 29
\bibitem[\protect\citeauthoryear{Santoro et~al.}{2016}]{santoro2016meta}
\begin{bchapter}
\bauthor{\bsnm{Santoro}, \binits{A.}},
\bauthor{\bsnm{Bartunov}, \binits{S.}},
\bauthor{\bsnm{Botvinick}, \binits{M.}},
\bauthor{\bsnm{Wierstra}, \binits{D.}},
\bauthor{\bsnm{Lillicrap}, \binits{T.}}:
\bctitle{Meta-learning with memory-augmented neural networks}.
In: \bbtitle{International Conference on Machine Learning},
pp. \bfpage{1842}--\blpage{1850}
(\byear{2016}).
\bcomment{PMLR}
\end{bchapter}
\endbibitem

%%% 30
\bibitem[\protect\citeauthoryear{Richards et~al.}{2021}]{richards2021adaptive}
\begin{bchapter}
\bauthor{\bsnm{Richards}, \binits{S.}},
\bauthor{\bsnm{Azizan}, \binits{N.}},
\bauthor{\bsnm{Slotine}, \binits{J.-J.}},
\bauthor{\bsnm{Pavone}, \binits{M.}}:
\bctitle{Adaptive-control-oriented meta-learning for nonlinear systems}.
In: \bbtitle{Robotics Science and Systems}
(\byear{2021})
\end{bchapter}
\endbibitem

%%% 31
\bibitem[\protect\citeauthoryear{Li and Xu}{2020}]{li2020unmanned}
\begin{bchapter}
\bauthor{\bsnm{Li}, \binits{Q.}},
\bauthor{\bsnm{Xu}, \binits{Y.}}:
\bctitle{Unmanned aerial vehicle angular velocity control via reinforcement
  learning in dimension reduced search spaces}.
In: \bbtitle{2020 American Control Conference (ACC)},
pp. \bfpage{4926}--\blpage{4931}
(\byear{2020}).
\bcomment{IEEE}
\end{bchapter}
\endbibitem

%%% 32
\bibitem[\protect\citeauthoryear{Fei et~al.}{2020}]{fei2020learn}
\begin{bchapter}
\bauthor{\bsnm{Fei}, \binits{F.}},
\bauthor{\bsnm{Tu}, \binits{Z.}},
\bauthor{\bsnm{Xu}, \binits{D.}},
\bauthor{\bsnm{Deng}, \binits{X.}}:
\bctitle{Learn-to-recover: Retrofitting uavs with reinforcement
  learning-assisted flight control under cyber-physical attacks}.
In: \bbtitle{2020 IEEE International Conference on Robotics and Automation
  (ICRA)},
pp. \bfpage{7358}--\blpage{7364}
(\byear{2020}).
\bcomment{IEEE}
\end{bchapter}
\endbibitem

%%% 33
\bibitem[\protect\citeauthoryear{Pi et~al.}{2021}]{pi2021general}
\begin{barticle}
\bauthor{\bsnm{Pi}, \binits{C.-H.}},
\bauthor{\bsnm{Dai}, \binits{Y.-W.}},
\bauthor{\bsnm{Hu}, \binits{K.-C.}},
\bauthor{\bsnm{Cheng}, \binits{S.}}:
\batitle{General purpose low-level reinforcement learning control for
  multi-axis rotor aerial vehicles}.
\bjtitle{Sensors}
\bvolume{21}(\bissue{13}),
\bfpage{4560}
(\byear{2021})
\end{barticle}
\endbibitem

%%% 34
\bibitem[\protect\citeauthoryear{Shi et~al.}{2019}]{shi2019neural}
\begin{bchapter}
\bauthor{\bsnm{Shi}, \binits{G.}},
\bauthor{\bsnm{Shi}, \binits{X.}},
\bauthor{\bsnm{O’Connell}, \binits{M.}},
\bauthor{\bsnm{Yu}, \binits{R.}},
\bauthor{\bsnm{Azizzadenesheli}, \binits{K.}},
\bauthor{\bsnm{Anandkumar}, \binits{A.}},
\bauthor{\bsnm{Yue}, \binits{Y.}},
\bauthor{\bsnm{Chung}, \binits{S.-J.}}:
\bctitle{Neural lander: Stable drone landing control using learned dynamics}.
In: \bbtitle{2019 International Conference on Robotics and Automation (ICRA)},
pp. \bfpage{9784}--\blpage{9790}
(\byear{2019}).
\bcomment{IEEE}
\end{bchapter}
\endbibitem

%%% 35
\bibitem[\protect\citeauthoryear{Ahmed et~al.}{2022}]{ahmed2022high}
\begin{bchapter}
\bauthor{\bsnm{Ahmed}, \binits{I.}},
\bauthor{\bsnm{Quinones-Grueiro}, \binits{M.}},
\bauthor{\bsnm{Biswas}, \binits{G.}}:
\bctitle{A high-fidelity simulation test-bed for fault-tolerant octo-rotor
  control using reinforcement learning}.
In: \bbtitle{2022 IEEE/AIAA 41st Digital Avionics Systems Conference (DASC)},
pp. \bfpage{1}--\blpage{10}
(\byear{2022}).
\bcomment{IEEE}
\end{bchapter}
\endbibitem

%%% 36
\bibitem[\protect\citeauthoryear{Schulman et~al.}{2017}]{schulman2017proximal}
\begin{botherref}
\oauthor{\bsnm{Schulman}, \binits{J.}},
\oauthor{\bsnm{Wolski}, \binits{F.}},
\oauthor{\bsnm{Dhariwal}, \binits{P.}},
\oauthor{\bsnm{Radford}, \binits{A.}},
\oauthor{\bsnm{Klimov}, \binits{O.}}:
Proximal policy optimization algorithms.
arXiv preprint arXiv:1707.06347
(2017)
\end{botherref}
\endbibitem

%%% 37
\bibitem[\protect\citeauthoryear{Sola and Sevilla}{1997}]{sola1997importance}
\begin{barticle}
\bauthor{\bsnm{Sola}, \binits{J.}},
\bauthor{\bsnm{Sevilla}, \binits{J.}}:
\batitle{Importance of input data normalization for the application of neural
  networks to complex industrial problems}.
\bjtitle{IEEE Transactions on nuclear science}
\bvolume{44}(\bissue{3}),
\bfpage{1464}--\blpage{1468}
(\byear{1997})
\end{barticle}
\endbibitem

%%% 38
\bibitem[\protect\citeauthoryear{Hutter et~al.}{2014}]{pmlr-v32-hutter14}
\begin{bchapter}
\bauthor{\bsnm{Hutter}, \binits{F.}},
\bauthor{\bsnm{Hoos}, \binits{H.}},
\bauthor{\bsnm{Leyton-Brown}, \binits{K.}}:
\bctitle{An efficient approach for assessing hyperparameter importance}.
In: \beditor{\bsnm{Xing}, \binits{E.P.}},
\beditor{\bsnm{Jebara}, \binits{T.}} (eds.)
\bbtitle{Proceedings of the 31st International Conference on Machine Learning}.
\bsertitle{Proceedings of Machine Learning Research},
vol. \bseriesno{32},
pp. \bfpage{754}--\blpage{762}.
\bpublisher{PMLR},
\blocation{Bejing, China}
(\byear{2014}).
\burl{https://proceedings.mlr.press/v32/hutter14.html}
\end{bchapter}
\endbibitem

%%% 39
\bibitem[\protect\citeauthoryear{Ahmed et~al.}{2018}]{ahmed2018comparison}
\begin{barticle}
\bauthor{\bsnm{Ahmed}, \binits{I.}},
\bauthor{\bsnm{Khorasgani}, \binits{H.}},
\bauthor{\bsnm{Biswas}, \binits{G.}}:
\batitle{Comparison of model predictive and reinforcement learning methods for
  fault tolerant control}.
\bjtitle{IFAC-PapersOnLine}
\bvolume{51}(\bissue{24}),
\bfpage{233}--\blpage{240}
(\byear{2018})
\end{barticle}
\endbibitem

%%% 40
\bibitem[\protect\citeauthoryear{Akiba et~al.}{2019}]{optuna_2019}
\begin{bchapter}
\bauthor{\bsnm{Akiba}, \binits{T.}},
\bauthor{\bsnm{Sano}, \binits{S.}},
\bauthor{\bsnm{Yanase}, \binits{T.}},
\bauthor{\bsnm{Ohta}, \binits{T.}},
\bauthor{\bsnm{Koyama}, \binits{M.}}:
\bctitle{Optuna: A next-generation hyperparameter optimization framework}.
In: \bbtitle{Proceedings of the 25th {ACM} {SIGKDD} International Conference on
  Knowledge Discovery and Data Mining}
(\byear{2019})
\end{bchapter}
\endbibitem

\end{thebibliography}
	%% if required, the content of .bbl file can be included here once bbl is generated
	%%\input sn-article.bbl

\end{document}